\documentclass[preprint,aps]{revtex4-1}
\usepackage{graphicx}
\usepackage{rotating}
\usepackage{amsmath}

\begin{document}
\title{Universality of Ultrasonic attenuation in amorphous systems at low temperatures}
\author{Pragya Shukla}
\affiliation{Department of Physics, Indian Institute of Technology, Kharagpur-721302, India }
\date{\today}

\widetext

\begin{abstract}

The competition between  unretarded dispersion interactions between molecules prevailing at  medium range order length scales    and their phonon induced coupling at larger scales  leads to appearance of nano-scale sub structures in amorphous systems. The complexity of intermolecular interactions gives rise to randomization of their operators. Based on a random matrix modelling of the Hamiltonian and its  linear response to an external strain field, we show that  the ultrasonic attenuation coefficient can be expressed as a ratio of two crucial length-scales related to molecular dynamics. A nearly constant value of the ratio for a wide range of materials then provides a theoretical explanation of the  experimentally observed qualitative universality of the ultrasonic attenuation coefficient at low temperatures.

\end{abstract}


\maketitle
.


\section{Introduction}

Experiments on thermal conductivity and acoustic attenuation in past have revealed an striking physical property  of amorphous systems at low temperatures i.e the universality of the ultrasonic attenuation coefficient $Q^{-1}$, defined as the ratio of wavelength $\lambda$ of the elastic wave to mean free path $l$ of the energy attenuation \cite{pohl} and also referred as {\it internal friction}. For  $T=0.1 \to 10 K$, $Q^{-1}(\omega; T)$   is found to be nearly independent of temperature $T$ as well as measuring frequency $\omega$. The magnitude of $Q^{-1}$ not only lies within about a factor of $20$ for all glasses but is also very small (around $\sim  10^{-4}$), indicating long (short) mean free paths at small (large) phonon frequencies.

Previous attempts to explain this behaviour  were based  on an assumed existence of the  defects modeled as tunnelling two level systems (TTLS) \cite{jack}. Although successful in explaining many  glass anomalies, the original TTLS model \cite{and, phil} suffered many drawbacks \cite{yl, lg1, galp}  (besides experimental lack of evidence supporting their existence  in most glasses). This encouraged attempts for improvements of the model by incorporating a phonon-TTLS interaction \cite{burin}, presence of TLS alongwith quasi- harmonic oscillators \cite{paras}  as well as  considerations of several new theories; (extensive research on this topic during previous decades renders it impossible to list all but a few leading to new theoretical developments  e.g. \cite{and, phil, galp, vlw, paras, yl, misha, ssp, lg1, spm, buch, gure, grig, schi2, emt, degi}).

 In context of the acoustic attenuation, an important direction  was taken in a  recent theory of coupled generic blocks with a phonon-mediated  interaction of type $1/r^3$ with $r$ as the separation between blocks \cite{vl, lg1}. A renormalisation approach used in \cite{vl} rendered the information regarding  the behavior  of a single generic  block unnecessary 
and provided useful insights  regarding  the universality at macroscopic scales. Although the  theory was later on  applied successfully to explain another glass-universality, namely,  Meissner-Berret ratio \cite{dl}, it has still left many  questions unanswered e.g how the block type structure appears, what is the effect of the intra-block forces over the inter-block ones, whether the universality is an emergent phenomenon occurring only at large scales or it also occurs at microscopic scales i.e for a single block; (for example, the study \cite{vl} does not provide any information about the attenuation coefficient for a  basic block). An answer to these questions is pertinent to understand the physical origin of universalities which motivates the present work.

Based on the nature of chemical bonding, the physics of solids is expected to vary at microscopic length scales. Contrary to other low temperature properties, however the ratio $\lambda/l$ is found to be universal  not only for glasses (with only few exceptions in some thin films) but for a huge class of materials different from them at large length scales e.g. disordered crystals, poly-crystals, some quasi crystals etc \cite{pohl}. Furthermore the irradiation experiments on crystalline silicon  for a wide range of radiation doses indicate the sound properties of the irradiated samples similar to glasses.   Another universality not confined only to glasses but applicable to many liquids too is that of excess vibrational density of states which can not be explained based on the phonon contributions only \cite{vdos}. These  universalities therefore seem to originate from more fundamental considerations, shared by both amorphous as well as disordered crystalline materials, with lack of long-range order not the cause of the low-energy excitations, and applicable not only for macroscopic sizes but also at microscopic scales (e.g. see \cite{chum} also in this context). This motivated us in \cite{bb1} to  consider the intermolecular interactions, more specifically Vanderwaals (VW) forces among the molecules within a block as the basis for the behavior;  it is important to emphasise here that VW forces among molecules are always present in all condensed phases and therefore are the natural candidates to decipher the experimentally observed universality.

Our primary focus in the present work is to seek the physical origin of the weak attenuation of the sound waves in amorphous systems.
For this purpose, it is necessary to first identify the local structures which respond to an external strain field by collective vibrations of molecules. But  phonons in a perfect harmonic dielectric crystal  are free of interactions, leading to a sound wave travel unattenuated. To understand long mean free paths in glasses at low frequencies, this intuitively suggest to seek for ordered structure, at least locally,  and repeated  almost periodically. The  structure  related to medium range order (MRO) in glasses seem to be playing the relevant role. (Note the glasses also have short range order but that is governed by covalent bonds which are quite rigid to undergo deformation by a weak strain field. Further the role of the molecular clusters or structural correlations  was proposed in past too e.g in \cite{vdos,gg1,yu1,du}) but it was not very well-defined \cite{ell3}). As discussed in \cite{bb3},  the size of the basic block  indeed turns out to be that of the length scale associated with medium range order. (Note the peculiarity of role played by MRO in context of acoustic modes was mentioned in \cite{ell3, mg} too; the study \cite{mg} indicated that  the continuum approximation for the medium, necessary for Debye formulation, breaks down   for acoustic modes with wavelength less than MRO). The combination of many such blocks can then provide required periodicity and their long-range interaction result in attenuation only at long length scales. Our theory of coupled blocks is therefore based on two main types of interactions, dominant at different spatial scales; a competition between them governs the block-size and also gives rise to an inter-connected block structure, with phonon mediated coupling of their stress fields. This in turn  leads to formulation of the attenuation in terms of the stress-stress correlations among basic blocks and their density of the states. As discussed later, both of them can be expressed in terms of the molecular properties which finally leads to a constant, system-independent average value of $Q^{-1}$.

The paper is organized as follows.  
The theory of an amorphous system of macroscopic size as a collection of  sub-structures coupled with each other via an inverse-cube phonon mediated interaction is discussed in detail  in \cite{vl,dl}; this is briefly reviewed in section II, with macroscopic solid referred as the super block and  the sub-structures referred as the basic blocks. 
Note the present work differs from \cite{dl, vl}  in context of the basic blocks details; the latter appear, in our theory, as a  result of VW interactions among molecules prevailing at nano-scales \cite{bb1}. The  theory  is used in section III to relate the $\langle Q^{-1} \rangle$ of the super block to the stress-stress correlations of the basic blocks, their  bulk density of states $\rho_e$ and volume $\Omega_b$;  here $\langle \rangle$ refers to the ensemble as well as spectral average. $\rho_e$ depends on a parameter $b$, referred as "bulk spectral parameter'' and derived in \cite{bb1} in terms of the molecular parameters. This along with $\Omega_b$ leads to dependence of $Q^{-1}$ on the length-ratio $R_0/ R_v$ with $R_0$ as the linear size of the basic block and $R_v$ as the distance between two nearest neighbor molecules mutually interacting by unretarded dispersion forces. A theoretical analysis of the ratio $R_0/R_v$ has indicated it to be a system-independent constant for amorphous systems (supported by data based on 18 glasses) \cite{bb3}.
In section IV, we express  $\langle Q^{-1}\rangle$ in terms of $R_0/ R_v$ and thereby theoretically prove its quantitative universality. It can however  be 
calculated directly from the molecular properties too; as discussed in section V, a good agreement of the results so obtained for 18 non-metallic glasses with experimental values not only lends credence to our theory of blocks but also provides an indirect route to reconfirm the universality of the ratio $R_v/R_0$. Note the 18 glasses chosen for comparison here are same as those used in \cite{bm}. A discussion of physical insights provided by our approach, brief comparison with other theories and approximations are outlined in section VI. We conclude in section VII with a summary of our main ideas and results.

\section{Super block: phonon mediated coupling of basic blocks}

The order at atomic dimensions in an amorphous solid is system dependent; it is sensitive to the nature of chemical bonding. 
The intuition suggests the universal properties to originate from the interactions which appear at length scales at which the solid manifests no system-dependence.  It is therefore relevant to seek and identify the  sub-units in the super block structure which  give rise to such interactions.  
For this purpose, let us first  express the Hamiltonian $H$ of the amorphous solid of volume $\Omega$ as the sum over intra-molecular interactions as well as inter-molecular ones
\begin{eqnarray}
H= \sum_k h_k({\bf r}_k) + {1 \over 2} \sum_{k,l}  {\mathcal U}(|{\bf r}_k - {\bf r}_l |)
\label{hat0}
\end{eqnarray}
with $h_k$ as the Hamiltonian of the $k^{th}$ molecule  at position ${\bf r}_k$
and ${\mathcal U}$ as an inter-molecular interaction with arbitrary range $r_0$. Assuming that all the relevant many body states are "localized", in the sense that the probability density for finding a given  molecule $"k"$ is "concentrated" (as defined by its mean square radius) in a region of finite radius $l$ around some point ${\bf r}_k$, it is possible to define a $3 D$ lattice (grid of points) ${\bf R}_{\alpha}$ with spacing $d \gg r_0$ such that the  molecule "k" is associated with  that lattice point    ${\bf R}_{\alpha}$  which is closest to ${\bf R}_{k}$. The association is fixed, is  insensitive to the dynamics and corresponds to representation of the solid by 3-dimensional  blocks of linear size $R_0$, with their centers at lattice points ${\bf R}_{\alpha}$.   The Hamiltonian $H$ can then be reorganised as a sum over basic block Hamiltonians and  the interactions between molecules on different blocks
\begin{eqnarray}
H= \sum_s {\mathcal H}^{(s)} + {1 \over 2} \sum_{s,t} \sum_{k \in s, l \in t}  {\mathcal U}(|{\bf r}_k - {\bf r}_l |)
\label{hat1}
\end{eqnarray}
where  ${\mathcal H}^{(s)}$ is the Hamiltonian  of a basic block labeled $''s''$, basically sum over the molecular interactions within the block : $ {\mathcal H}^{(s)} = \sum_{k \in s} h_k({\bf r}_k) + {1 \over 2} \sum_{k,l \in s}  {\mathcal U}(|{\bf r}_k - {\bf r}_l |)$. As mentioned below, the molecules interactions appearing in 2nd term in eq.(\ref{hat1}) rearrange themselves collectively and results in emergence of coupled stress fields of the blocks. The number $g$ and volume $\Omega_b$ of these blocks can be determined by analysing the competition between inter-molecular forces with emerging forces i.e phonon mediated coupling: $g=\Omega/\Omega_b$ with $\Omega_b \sim R_0^3$.  The statistical behavior of the Hamiltonian ${\mathcal H}$   is discussed in detail in \cite{bb1}.

To analyze the ultrasonic attenuation in glasses, we first need to analyze the response of ${\mathcal H}$ to an external strain field.

\subsection{Perturbed Hamiltonian of a basic block}

In presence of an external strain field, the molecules in a glass block are displaced from their equilibrium position and their interactions with those in surrounding blocks give rise to  a stress field distributed over the block. Let $u({\bf r})$ be the displacement, relative to some arbitrary reference frame, of the matter at point ${\bf r}$, the elastic strain tensor  can then be defined as 

\begin{eqnarray}
e_{\alpha \beta }({\bf r},t) &=&  {1 \over 2} \left(  {\partial  u_{\alpha} \over \partial x_{\beta}} +  {\partial  u_{\beta} \over \partial x_{\alpha}}   \right)
\label{str}  
\end{eqnarray}
with subscripts $\alpha, \beta$ referring to the tensor-components. 

This gives rise to stress in the block which can in general have both elastic as well as inelastic components. The perturbed Hamiltonian $H_{pt}$ of the basic block, labeled $''s''$ can then be written as a sum over elastic and inelastic  contributions 

\begin{eqnarray}
{ \mathcal H}_{pt}^{(s)} =  {\mathcal H}_{pt,ph}^{(s)} + {\mathcal H}_{pt,nph}^{(s)}.
\end{eqnarray} 

Each of these parts can further be expanded as a Taylor' series around unperturbed block Hamiltonian ${\mathcal H}_x$ in terms of strain $e_{\alpha \beta}$ in long wavelength limit (where  the subscript $"x''$  refers to the elastic ($''x=ph''$) and inelastic parts ($''x=nph''$) respectively):

\begin{eqnarray}
{\mathcal H}^{(s)}_{pt, x}(t) = {\mathcal H}_{x}^{(s)} + \int {\rm d}{\bf r} \; e_{\alpha \beta }({\bf r},t) \; {\Gamma}^{(s)}_{\alpha \beta; x}({\bf r}) + O(e_{\alpha \beta }^2) 
\label{hpt1}
\end{eqnarray}

with  $\Gamma^{(s)}_{\alpha \beta; x}({\bf r})$ as the stress tensor; as clear from above 
$\Gamma^{(s)}_{\alpha \beta; x}({\bf r})  =  {\partial {\mathcal H}^{(s)}_{pt, x} \over \partial e_{\alpha \beta} } $. Further, assuming the isotropy and the small block-size, the distributed stress field within the block of volume $\Omega_b$ can be replaced by an average acting from the centre of mass of the block: $\int_{\Omega_b} \; {\rm d}{\bf r} \; \Gamma^{(s)}_{\alpha \beta} ({\bf r}) = \Gamma^{(s)}_{\alpha \beta} $.  The perturbed Hamiltonian of the basic block can then be approximated as
 
\begin{eqnarray}
{ \mathcal H}_{pt;x}^{(s)} = {\mathcal H}^{(s)}_x +  \sum_{\alpha \beta} e^{(s)}_{\alpha \beta}  \; {\Gamma}^{(s)}_{\alpha \beta; x} 
\label{ha2}
\end{eqnarray}
with $e^{(s)}_{\alpha \beta} (t)$ referring to the phonon strain field $e_{\alpha \beta} ({\bf r}, t)$ at the $s$-th block.

\subsection{Super block Hamiltonian}

The super block consists of $g$ basic blocks, perturbed by mutual interaction. 
To proceed further, it is  useful to separate its Hamiltonian $H$ into phononic and non-phonon contributions (referred by subscripts $''ph''$ and $''nph''$ respectively): $H= H_{ph} + H_{nph}$ (\cite{lg1}). The contribution of elastic part $H_{ph}$ to the ultrasonic attenuation 
in glass super block at temperatures $T < 1^o K$   is negligible. We therefore need to consider the contribution  from the  inelastic part $ H_{ nph}$  only; to reduce notational complexity,  henceforth, the subscripts $''nph''$   will be suppressed and the notations $H, {\mathcal H}^{(s)}_{pt}, \Gamma^{(s)}$ etc will be used for $H_{nph}, {\mathcal H}^{(s)}_{pt;nph}, \Gamma^{(s)}_{nph}$ respectively. 

As the strain tensor $e_{\alpha \beta}$ contains a contribution from the phonon field, the exchange of virtual phonons will give rise to an effective (“RKKY”-type) coupling between the stress tensors of any two block-pairs. Let $\Gamma^{(s)}_{\gamma \delta} ({\bf r})$ be the stress tensor at point ${\bf r}$ of the basic block ''s''. The interaction $V_{st}$ between the blocks "s" and "t " can be given as \cite{vl}

\begin{eqnarray}
V_{st}  &=& {1\over 4 \pi \rho_m c^2} \; \int_s {\rm d}{\bf r} \; \int_{t} {\rm d}{\bf r'}  \; \sum_{te} \;  
{\kappa^{(st)}_{\alpha \beta \gamma \delta}  \over   | \; {\bf r}-{\bf r'} \; |^3 }. 
 \; \;   \Gamma^{(s)}_{\alpha \beta} ({\bf r})\otimes
\;  \Gamma^{(t)}_{\gamma \delta} ({\bf r'}) 
\label{vsb}
\end{eqnarray}

with $\rho_m$ as the mass-density and $c$ as the speed of sound in the super block. Here the subscripts $\alpha \beta \gamma \delta$ refer to the tensor components
and the symbol $\sum_{te}$ refers to a sum over all tensor components: 
$\sum_{te} \equiv \sum_{\alpha \beta \gamma \delta}$. 
The directional dependence of the interaction is represented by $\kappa^{(st)}_{\alpha \beta \gamma \delta}=\kappa^{(st)}(\theta, \phi)$; it is assumed to depend only on the relative orientation ($\theta, \phi$) of the block-pairs and is  independent from their relative separation \cite{dl}:

\begin{eqnarray}
\kappa^{(st)}_{ijkl} &=& - \left(\delta_{jl} - 3 n_j n_l \right)  \delta_{ik}   
+ \nu_2 \;\left[ - (\delta_{ij} \delta_{kl} + \delta_{ik} \delta_{jl} + \delta_{il} \delta_{jk}) + \right . \nonumber \\
&+& \left.  3 \; \left(n_j n_l \delta_{ik} + n_j n_k \delta_{il} + n_i n_k \delta_{jl} 
   + n_i n_l \delta_{jk} +  n_i n_j \delta_{kl} + n_k n_l \delta_{ij} \right)  
- 15 \; \sum_{ijkl} \; n_i n_j n_k n_l  \right]
\label{ad3}
\end{eqnarray}
where $\nu_2= \left(1-{c_t^2 \over c_l^2}\right)$ and ${\bf n} = n_1 \hat i + n_2 \hat j + n_3 \hat k$ is the unit vector along the direction of position vector ${\bf r-r'}$. 
Again assuming the isotropy and the small block-size, the interaction between various points of the block-pairs can be replaced by the average interaction between their centers ${\bf R_s}$ and ${\bf R_t}$. The phonon mediate coupling between the blocks can then be  approximated as  \cite{vl, dl}

\begin{eqnarray}
V_{st}  &=&  {1\over 4 \pi \rho c^2} \; 
\sum_{\alpha \beta \gamma \delta} \;  
{ \kappa^{(st)}_{\alpha \beta \gamma \delta} \over   | \; {\bf R_s}-{\bf R_{t}} \; |^{3} }  \; \;   \Gamma^{(s)}_{\alpha \beta} \otimes
\;  \Gamma^{(t)}_{\gamma \delta}  
\label{zq++}
\end{eqnarray}

Due to the above emerging interactions at large length scales, the super block Hamiltonian in eq.(\ref{hat1}) is not just a sum over basic block Hamiltonians but also includes their phonon mediated coupling.

Eq.(\ref{zq++}) describes an emerging interaction at large length scales. The Hamiltonian of the  super block  in eq.(\ref{hat1}) can now be rearranged as a sum over those of the basic blocks as well as  their phonon mediated coupling. 
In absence of external strain field, the non-phonon part of $H$ can be rewritten as  
\begin{eqnarray}
H=H_0 +V
\label{zq+} 
\end{eqnarray}
with $H_0$ as a sum over non-phonon part of the unperturbed basic block Hamiltonians, $H_0 = \sum_{s=1}^g \;   {\mathcal H}^{(s)}$, and, $V$ as the net pair-wise interaction among blocks: $V=\sum_{s,t; s\not=t} V_{st}$ where $\sum_{s,t}$ implies the sum over all basic blocks. 
The presence of a weak external strain field perturbs the basic blocks and thereby $H$. The non-phonon part of the perturbed Hamiltonian $H_{pt}$  can  be written as \cite{vl,dl}
\begin{eqnarray}
H_{pt} = H + \sum_{s=1}^g \sum_{\alpha \beta} e_{\alpha \beta}^{(s)}  \; \Gamma_{\alpha \beta}^{(s)} 
= H + \sum_{\alpha \beta} e_{\alpha \beta} \; \Gamma_{\alpha \beta}
\label{zq}
\end{eqnarray}
where the 2nd equality follows by assuming the same strain operator for all blocks $ e_{\alpha \beta}^{(s)} \approx e_{\alpha \beta}$ and writing  $\Gamma_{\alpha \beta} = \sum_{s=1}^g \Gamma_{\alpha \beta}^{(s)}$.
(Note, as discussed in \cite{dl}, the total Hamiltonian for the super block contains two additional terms besides $V$ (see eq.(2.21) in \cite{dl}) but their ensemble averaged contribution is negligible. Alternatively it can also be absorbed by redefining stress operators).

\section{Ultrasonic attenuation coefficient: relation with stress matrix}

The dimensionless ultrasonic attenuation coefficient or internal friction $Q_{a}^{-1}(\omega)$ of a phonon of frequency $\omega$  and wavelength $\lambda$  can in general be defined as \cite{dl, pohl}

\begin{eqnarray}
 Q_{a}^{-1} ={1 \over 2 \pi^2 } \; {\lambda \over l} 
\label{Q0}
\end{eqnarray}
with $l$ as its mean free path. Note the above definition is different from that in \cite{pohl} by a constant: $Q_{a,pohl}^{-1} = \pi \; Q_{a}^{-1}$. 

Consider the attenuation of acoustic waves in a glass super block with its Hamiltonian $H$ given by eq.(\ref{zq}). 
Assuming the coupling between phonon and non-phonon degrees of freedom a weak perturbation on the phonon dynamics, $ Q_{a}^{-1}(\omega)$ can be expressed as \cite{dl}

\begin{eqnarray}
 Q_{a}^{-1}(\omega) = ( \pi \; \rho_m  \; c_a^2)^{-1} \; {\rm Im} \; \chi_a(\omega)
\label{Q}
  \end{eqnarray}
with $\rho_m$ as the mass-density of the material, $c_a$ as the speed of acoustic wave in the longitudinal (with $a =l$) or transverse direction ($a=t$). Here $\chi_{l,t}(\omega)$, referred as the longitudinal or transverse response function, are the measures of the linear response of the basic blocks to external strain field and can be defined as follows.

\subsection{Non-phonon linear response function}

Consider the linear response of a basic block, labeled as $''s''$, to an external strain field $e_{ij}({\bf r}, t)= e_{ij} \; exp[i({\bf q.r} - \omega t)]$ with $e_{ij}$ real but infinitesimal. The perturbed Hamiltonian is given by eq.(\ref{ha2})  with corresponding  stress-field given as 
$\Gamma^{(s)}_{ij}({\bf r}, t) = \langle \Gamma^{(s)}_{ij} \rangle \; exp[i({\bf q.r} - \omega t)]$ where $\langle \Gamma^{(s)}_{ij}\rangle$ is in general complex. 

The complex response function or the susceptibility for a basic block can then be defined as 
\begin{eqnarray}
\chi^{(s)}_{\alpha \beta \gamma \delta}({\bf q}, \omega) \equiv  \frac{1}{\Omega_b} \;  {\partial \langle \Gamma^{(s)}_{\alpha \beta} ({\bf q}, \omega) \rangle \over \partial e_{\gamma \delta}}.
\label{chi-}
\end{eqnarray}
Here in general the variable ${\bf q}$ and $\omega$ are independent variables. But as our interest is in values of $q$ close to $\omega/c_{l,t}$ (with $c_{l,t}$ as the longitudinal and transverse speeds of sound in the glass solid), $\chi$ will henceforth be written as a function of $\omega$ only \cite{vl}.

 The  imaginary part of $\chi^{(s)}(\omega)$ can be written  in the representation in which unperturbed basic block Hamiltonian ${\mathcal H}^{(s)}$ is diagonal (later referred as non-interacting or NI basis). Let $|{m_s} \rangle$, $m_s =  1\rightarrow N$ be the many body eigenstate of ${\mathcal H}^{(s)}$ with energy $e_m$, then

\begin{eqnarray}
{\rm Im} \; \chi^{(s)}_{\alpha \beta \gamma \delta} (\omega) &=& {(1-{\rm e}^{-\beta \omega})\over Z}\;  \sum_m \; e^{-\beta e_m} \;  \chi^{(m,s)}_{\alpha \beta;\gamma \delta}(\omega)    
\label{chi0}
\end{eqnarray}
with Z as the partition function.  Here to simplify presentation, we set $\hbar=1$. Further
\begin{eqnarray}
 \chi^{(m,s)}_{\alpha \beta \gamma \delta}(\omega) =
\frac{\pi }{\Omega_b}  \; 
\sum_{n=1}^N \; \Gamma^{(s)}_{\alpha \beta; mn}  \;  \Gamma^{(s)}_{\gamma \delta; n m} \;\; \delta(e_n-e_m - \omega) 
\label{chi1}
\end{eqnarray}
with $\Gamma^{(s)}_{\alpha \beta; kl}$ as the matrix element of the stress-tensor in the NI basis:
$\Gamma^{(s)}_{\alpha \beta; kl} = \langle {k_s} | \; \Gamma^{(s)}_{\alpha \beta} \; | {l_s} \rangle $. 
In general $\chi^{(m)}_{\alpha \beta \gamma \delta}$ depends on the energy level $e_m$ and fluctuates  over the spectrum. It is then useful to define the spectral averaged susceptibility over the $N$-level spectrum of the basic block 

\begin{eqnarray}
{\langle\chi^{(s)}_{\alpha \beta \gamma \delta} \rangle_{\omega}} &=& {1\over N \omega_c} \sum_{m=1}^N  \int_0^{\omega_c}  \;    \chi^{(m,s)}_{\alpha \beta \gamma \delta}(\omega-e_m)  \; {\rm d}\omega  
\label{chia}
\end{eqnarray}
where $\omega_c$ is the bulk spectrum width of the basic block.

Furthermore the fluctuations of $\Gamma^{(s)}_{\alpha \beta; kl}$ as well as those of the energy levels over the ensemble also  influence $\chi^{(m,s)}_{\alpha \beta \gamma \delta}(\omega) $ and it is appropriate to consider its ensemble average $ \langle \chi^{(m,s)}_{\alpha \beta \gamma \delta}(\omega) \rangle_e$ too.   Assuming  isotropy, rotationally invariance of  the  basic block (as its linear size $L \gg a$ with $a$ as the atomic length scale), all $3^8$ components of response function can further be expressed in terms of the transverse and longitudinal response \cite{dl}:

\begin{eqnarray}
\langle \chi^{(s)}_{\alpha \beta \gamma \delta}(\omega) \rangle_{e, \omega}  
&=&  \left( q_c \; \delta_{\alpha \beta} \delta_{\gamma \delta} +
\delta_{\alpha \gamma} \delta_{\beta \delta}  + \delta_{\alpha \delta} \delta_{\beta \gamma}  \right) \; \langle \chi^{(s)}_t \rangle_{e, \omega}
\label{chim}
\end{eqnarray}
where $q_c={\langle \chi^{(s)}_l\rangle_{e,\omega} \over \langle\chi^{(s)}_t\rangle_{e,\omega}} - 2$ along with $\langle . \rangle_e$ implying an ensemble averaging, $\langle . \rangle_{e, \omega}$  an averaging over both $\omega$ and ensemble.

The relations given in eq.(\ref{chi-}) to eq.(\ref{chim}) are applicable for a basic block of volume $\omega_b$. Following similar forms of eq.(\ref{ha2}) and eq.(\ref{zq}), these can be generalized for the susceptibility 
$\langle \chi_a\rangle^{sup}_{e,\omega}$ of a super block.
This follows by dropping the superscript $"s"$ and with replacements  $\Omega_b \to \Omega$, $N \to N^g, \omega_c \to W_c, e_n \to E_n$ in eq.(\ref{chi-}) to eq.(\ref{chim}); note here $E_m$ refers to a many body energy level of $H$ (defined in eq.(\ref{zq+}).

\subsection{Relation between  $Q_{a}^{-1}$ and stress-correlations }

{\noindent {\bf For Basic Block:}} Due to disorder beyond atomic 
scales, a typical  matrix element of the stress tensor of a basic block   fluctuates over the 
ensemble and can be both positive as well as negative. This implies $\langle \Gamma^{(s)}_{\alpha \beta; kl} \rangle_e  =0$.
Further, at temperature $T =0$,  the spectral averaging (defined in eq.(\ref{chia})) of  eq.(\ref{chi1}) 
followed by an ensemble averaging leads to the stress-stress correlation of the basic block

\begin{eqnarray}
\sum_{m, n=1}^N 
\langle
\Gamma^{(s)}_{\alpha \beta; mn}  \;  \Gamma^{(s)}_{\gamma \delta; n m } \rangle_{e}
={ N \omega_c \;  \Omega_b \over \pi} 
 \left( q_c \; \delta_{\alpha \beta} \delta_{\gamma \delta} +
\delta_{\alpha \gamma} \delta_{\beta \delta}  + \delta_{\alpha \delta} \delta_{\beta \gamma}  \right) \; \langle {\rm Im} \; \chi^{(s)}_t \rangle_{e.\omega}
\label{stp1}
\end{eqnarray}
where $\langle {\rm Im} \; \chi^{(s)}_t \rangle_{e.\omega}$ is defined in eq.(\ref{chim}).

The short-range order of  atomic positions in the basic-block along with its small size suggests 
a homogeneous nature of  many body interactions. 
The ensemble averaged matrix elements of $\Gamma^{(s)}_{\alpha \beta}$, in the NI basis i.e 
the eigenfunction basis of $H^{(s)}_0$,  can then be assumed to be of almost same 
strength. ({\it This is equivalent to say that, due to small size of block, stress can be 
assumed to be homogeneous i.e of the same order everywhere in the block. This assumption 
therefore puts a constraint on our basic-block size}).  One can then write $\sum_{ m, n=1}   \langle \Gamma^{(s)}_{\alpha \beta; mn}  \;  \Gamma^{(s)}_{\gamma \delta; n m }  \rangle_{e}
= N^2 \; \langle \Gamma^{(s)}_{\alpha \beta; mn}  \;  \Gamma^{(s)}_{\gamma \delta; n m }  \rangle_{e}$. This on substitution  in eq.(\ref{stp1})  leads to

\begin{eqnarray}
  \langle \Gamma^{(s)}_{\alpha \alpha; mn}  \;  \Gamma^{(s) }_{\gamma \gamma; mn}  \rangle_{e}
&=& { \omega_c \; \Omega_b  \over  N \pi}  \;  \left[q_c  + \delta_{\alpha \gamma} \right]  
\; \langle {\rm Im} \; \chi^{(s)}_t(\omega)  \rangle_{e, \omega} 
\label{gcorr1}   \\
\langle \Gamma^{(s)}_{\alpha \beta; mn}  \;  \Gamma^{(s) }_{\alpha \beta;  m n }  \rangle_{e}
&=& 
\langle \Gamma^{(s)}_{\alpha \beta; mn}  \;  \Gamma^{(s) }_{\beta \alpha; m n}  \rangle_{e}
= { \omega_c \; \Omega_b  \over  N \pi}  \;\langle {\rm Im} \; \chi^{(s)}_t(\omega)  \rangle_{e, \omega}   \hspace{0.5in} \alpha \not= \beta
\label{gcorr2}
\end{eqnarray}

Further using eq.(\ref{Q}) in eqs.(\ref{gcorr1}, \ref{gcorr2}),  the correlations can be expressed  in terms of the average ultrasonic absorption $\langle Q^{-1}_t(\omega)  \rangle_{e,\omega}$ of the basic block

\begin{eqnarray}
  \langle \Gamma^{(s)}_{\alpha \alpha; mn}  \;  \Gamma^{(s) }_{\gamma \gamma; mn}  \rangle_{e}
&=& {N^{-1} \; \omega_c \; \rho_m \; c_t^2 \; \Omega_b  }  
\; \langle Q^{-1}_t(\omega)  \rangle_{e,\omega} \; \delta_{\alpha \gamma}
\label{gcorr3}   \\
\langle \Gamma^{(s)}_{\alpha \beta; mn}  \;  \Gamma^{(s) }_{\alpha \beta;  m n }  \rangle_{e}
&=& 
\langle \Gamma^{(s)}_{\alpha \beta; mn}  \;  \Gamma^{(s) }_{\beta \alpha; mn}  \rangle_{e}
= {N^{-1} \; \omega_c \; \rho_m \; c_a^2 \; \Omega_b}  \langle Q^{-1}_t(\omega)  \rangle_{e,\omega}  
\label{gcorr4}
\end{eqnarray}

Eq.(\ref{gcorr4})  can be rewritten in terms of the mean-square matrix element $\nu^2 = \langle \left(\Gamma^{(s)}_{\alpha \beta; mn} \right)^2 \rangle_{e}$ 
\begin{eqnarray}
\langle  Q_{a}^{-1} \rangle_{e, \omega}
&=&    { N \; \nu^2 \over    \omega_c \;  \rho_m \; c_a^2 \; \Omega_b } 
=  { \gamma^2  \over   \omega_c \; \rho_m \; c_a^2 \; \Omega_b }  
\label{gcorr5}
\end{eqnarray}
where  $\gamma^2  \equiv N^{-1} \; {\rm Tr} (\Gamma^{(s)}_{\alpha \beta})^2 = N \nu^2 $ is related to the coefficient of the phonon mediated coupling  $V$ between two basic blocks (which  is of the form ${\gamma^2\over 8 \pi \rho_m c^2 r^3}$, see  eq.(\ref{zq++})). 

As discussed in \cite{bb1}, the ensemble averaged density of the states which participate in these excitations, has a universal form  in the bulk of the spectrum: $\langle \rho_{bulk}(e) \rangle  =    {N  b \over  2\pi } \;  \sqrt{2 - \left({b e}\right)^2}$
with $b$ later referred as the bulk spectral parameter and $\langle \rangle$ as the ensemble average; (note here $\langle \rho_e(e) \rangle$ is normalised to $N$: $\int \langle \rho_e(e) \rangle \; {\rm d}e =N$). This gives the bulk spectral width as
\begin{eqnarray}
\omega_c = {2 \sqrt{2} \over b}  = {2 N \over \pi \langle \rho_{bulk}(0) \rangle} 
\label{delb}
\end{eqnarray}
As discussed in detail in \cite{bb1},  $b$ can be expressed as

\begin{eqnarray}
b  \approx   \; { 36  \over   \eta  \; \sqrt{ z \; g_0} \; A_H}  \; {y^6 \over (1+y)^6}
=  { 9  \over 4 \; \sqrt{ 3 } \; A_H }  \; \left({y \over 1+y}\right)^{9/2}
\label{b}
\end{eqnarray}  
with $A_H$ as the Hamaker constant  of the material, $z$ as the average number of nearest neighbors of a given molecule, $g_0$ as the number of molecules in the basic block, $\eta={\mathcal N}-1$ with ${\mathcal N}$ as the  number of relevant vibrational energy levels in a molecule) and $2 R_v$ as the distance between the centers of two nearest neighbor molecules. Based on the structural stability analysis of the amorphous systems, $z$ is predicted to be of the order of $3$ (for a three dimensional block) \cite{phil}.  Further    ${\mathcal N}$ corresponds to the number of single molecule states participating in dispersion interaction with another molecule. Alternatively, this is the number of dipole transitions among vibrational states of a molecule due to dispersion interaction with another one. Usually the allowed number of such transitions is 3 ($\delta m = 0, \pm 1$ with $m$ as the quantum number of the state); in any case weak nature of the dispersion interaction rules out higher number of such transitions). 
Note ${1 \over \omega_c \Omega_b}$ is of the order of  the bulk-density per unit volume. This in turn renders $\langle Q_{a}^{-1} \rangle_{e, \omega}$ given by eq.(\ref{gcorr5}) analogous to  that of  TTLS model: 
$ \langle Q_{a}^{-1} \rangle_{e, \omega}^{\rm TTLS}= {\pi  \; \gamma^2 \; {\overline P}  \over   2 \; \rho_m \; c_a^2 }  $ with ${\overline P}$ as the density of states of TTLS per unit volume.


\vspace{0.1in}

{\noindent {\bf For Super Block:}} eq.(\ref{gcorr5}) corresponds to the average coefficient of attenuation in a basic block. Proceeding exactly as above, the average coefficient for a super block, say $\langle  Q_{a}^{-1} \rangle^{sup}_{e, \omega}$, can also be obtained. The steps are as follows.  Eq.(\ref{stp1}) is now replaced by the relation
\begin{eqnarray}
\sum_{m, n=1}^{N^g} 
\langle
\Gamma_{\alpha \beta; mn}  \;  \Gamma_{\gamma \delta; n m } \rangle_{e}
={ N^g W_c \;  \Omega \over \pi} 
 \left( q_c \; \delta_{\alpha \beta} \delta_{\gamma \delta} +
\delta_{\alpha \gamma} \delta_{\beta \delta}  + \delta_{\alpha \delta} \delta_{\beta \gamma}  \right) \; \langle {\rm Im} \; \chi_t \rangle^{sup}_{e.\omega}
\label{stp2}
\end{eqnarray}
where $\Gamma_{\alpha \beta; mn}$ refers to the matrix element of $\Gamma_{\alpha \beta}$ in the eigenbasis of $H$ (eq.(\ref{zq+}). But noting that the left side of eq.(\ref{stp2}) can be rewritten as 
$\langle
{\rm Tr} (\Gamma_{\alpha \beta; mn})^2 \rangle$ and is therefore basis-invariant, it can be evaluated in the eigenbasis of $H_0$ i.e the product basis of single block states referred as $|E^0_n\rangle$, $n=1 \to N^g$. Using  
\begin{eqnarray}
\Gamma_{\alpha \beta; mn} = \sum_{s=1}^g \Gamma^{(s)}_{\alpha \beta; mn}
\label{su1}
\end{eqnarray}
 along with $\langle\Gamma^{(s)}_{\alpha \beta; mn}\Gamma^{t)}_{\alpha \beta; mn} \rangle =0$, it is easy to see that 
\begin{eqnarray}
\sum_{m, n=1}^{N^g} \langle
\Gamma_{\alpha \beta; mn}  \;  \Gamma_{\gamma \delta; n m } \rangle_{e}
= g \; N^{g+1}  \; \nu^2.
\label{su2}
\end{eqnarray}
The above follows because $\Gamma^{(s)}_{\alpha \beta; mn} \not= 0$ only if 
the product states $|E^0_m\rangle$ and $|E^0_n\rangle$ differ only by the contribution from the $s^{th}$ basic block. Further this also implies that  the relevant spectral averaging for the super block is same as that of a basic block i.e $W_c = w_c$. The above, along with the definition   $\langle  Q_{a}^{-1} \rangle^{sup}_{e, \omega} = ( \pi \; \rho_m  \; c_a^2)^{-1} \; \langle {\rm Im} \; \chi_a \rangle^{sup}_{e.\omega}$ and $\Omega=g \; \Omega_b$, now leads to

\begin{eqnarray}
\langle  Q_{a}^{-1} \rangle^{sup}_{e, \omega}
&=&    { N \; g \; \nu^2 \over    \omega_c \;  \rho_m \; c_a^2 \; \Omega } 
=  { \gamma^2  \over   \omega_c \; \rho_m \; c_a^2 \; \Omega_b }  
\label{gcorr5s}
\end{eqnarray}
 A comparison of the above result with eq.(\ref{gcorr5}) clearly indicates that
\begin{eqnarray}
\langle  Q_{a}^{-1} \rangle^{sup}_{e, \omega} =\langle  Q_{a}^{-1} \rangle_{e, \omega} 
\label{qsup}
\end{eqnarray}

\vspace{0.1in}

\section{Qualitative universality of $Q_{a}^{-1}$ }

Based on unretarded dispersion interaction between molecules, 
eq.(\ref{gcorr5}) relates the ultrasonic attenuation coefficient $\langle Q_a^{-1} \rangle$ 
to the bulk spectrum width $\omega_c$ and thereby bulk spectrum parameter $b$. Eq.(\ref{b}) expresses $b$  in terms of the molecular properties. Further  as discussed in \cite{bb3} in detail, the size $t$ of the basic block can be expressed as
\begin{eqnarray}
 t^2 =  { R_0^3\over 4 \; R_v}
 \label{ve0}
 \end{eqnarray}
 Here  $R_v$ is the distance of closest separation between two molecules in the material and  $R_0$ is a length scale at which dispersion interaction between two moelcules (i.e basic structural units) is balanced by the phonon mediated coupling of their stress fields \cite{bb3}
\begin{eqnarray}
R_0^3 = {\rho_m  \; c^2 \; C_6\over   8 \gamma_m^2}.
\label{r03}
\end{eqnarray}
with $C_6$ as the dispersion coefficient and $\gamma_m$ as the coupling strength of the stress fields of the molecules.  Using $\Omega_b = s \; t^3$, the above  then gives the  volume $\Omega_b$  of the basic block in terms of molecular parameters. Further, as discussed in \cite{bb3}, the number of molecules in a basic block can  be given as 

\begin{eqnarray}
g_0 =  {\Omega_b \over \Omega_{\rm eff}} \approx {1 \over (1+y)^3} \; \left({t \over R_m}\right)^3 =  {y^3 \over 8 \; (1+y)^3} \; \left({R_0 \over R_v}\right)^{9/2}   
\label{g0}
\end{eqnarray}

A combination  of the above relations then gives  $\langle Q_a^{-1} \rangle$  in terms of the molecular properties. This can be derived as follows.
A substitution of  eq.(\ref{b}) in eq.(\ref{gcorr5}), along with above relations for $t, R_0$ and $g_0$ and $s=4 \pi/3$,  leads to 

\begin{eqnarray}
 \langle  Q_{a}^{-1} \rangle_{e, \omega}
&\approx &  { 64 \; \gamma^2 \over    2 \; s \; \eta \sqrt{2 z g_0 } \; \rho_m \; c^2 \; C_6} \; {R_v^6 \over t_0^3 } 
\label{qb1}
\end{eqnarray}
where, as discussed in {\it appendix C}, $\gamma$, the coupling strength of basic blocks can be expressed in terms of that of molecules i.e $\gamma_m$,
\begin{eqnarray}
\gamma^2  \approx {4 \; \pi \; g_0 \over K \sqrt{2}} \; \gamma_m^2 
\label{gam}
\end{eqnarray}
with 
\begin{eqnarray}
K^2=18 \left(5-4 \; {c_t^2 \over c_l^2}\right).
\label{kk}
\end{eqnarray}

Using eq.(\ref{r03}) to replace $C_6$  in the above equation leads to

\begin{eqnarray}
 \langle  Q_{a}^{-1} \rangle_{e, \omega}
&\approx & { 8 \; \pi \; \sqrt{g_0}   \over  s \;  \eta \; \sqrt{z}  \; K} \; {R_v^6\over R_0^3 \; t^3} \\
&=& { 32 \; \pi \; \over  s \;  \eta \; \sqrt{2 z}  \; K} \; \left({y\over (1+y)}\right)^{3/2} \; \left({R_v\over R_0}\right)^{21/4}
\label{b4} 
\end{eqnarray}
Here the $2^{nd}$ equality is obtained by substitution of $t$ and $g_0$ from  eq.(\ref{ve0}) 
and eq.(\ref{g0}). Further, as mentioned below eq.(\ref{b}),  $\eta =2$ (with ${\mathcal N}=3$ as the number of allowed dipole transitions in a molecule) and $z$ as the number of nearest neighbors of a molecule  (those only interacting by VWD). The quantitative information about $R_v$  available for a wide range of materials suggests  $R_v \sim R_m$ (\cite{phil2}). Taking   $y= {R_v \over R_m} \sim 1$ leads to, from eq.(\ref{g0}), $g_0 \approx 8$. Assuming uniform mass density, this also implies only  three nearest neighbor molecules to a given molecule within a spherical basic block  of radius $t=\sqrt{ R_0^3\over 4 \; R_v}$ and therefore $z=3$.

Following from eq.(\ref{b4}), an almost quantitative universality  of $Q^{-1}$, as experimentally predicted for  amorphous systems \cite{pohl}, is not directly obvious. This
however follows by noting that the length scales $R_0$ and $R_v$ are related by a constant: $R_0=4 R_v$ \cite{bb3}.  Substitution of ${R_0\over R_v} =4$  in eq.(\ref{b4}) along with $y \approx 1$ and $ s=4 \pi /3$ leads to an almost material independent value  of average internal friction
\begin{eqnarray}
\langle  Q_{a}^{-1} \rangle_{e, \omega} \approx   2.83 \times 10^{-4}.  \; \left(1.25 -  {c_t^2 \over c_l^2}\right)^{-1/2}
\label{qth}
\end{eqnarray}
As discussed in \cite{bm, dl}, the ratio ${c_t \over c_l}$, and therefore $K$ (from eq.(\ref{kk})), is almost constant for many structural glass.
(Previous experiments indicate that ${c_l \over c_t}$ varies between $1.5 \to  2$, thus changing $\langle  Q_{a}^{-1} \rangle_{e, \omega}$ within $10\%$ only).

Further insight in the above result can be gained by rewriting
$\langle  Q_{a}^{-1} \rangle_{e, \omega} $ in terms  of the approximate number of molecules, say $g_0$,  in a basic block. Substitution of eq.(\ref{g0}) in  eq.(\ref{b4}) gives $\langle  Q_{a}^{-1} \rangle_{e, \omega} \propto {g_0^{-7/6}  }$. Further, using the relation  $R_0=4 R_v$ \cite{bb3}, eq.(\ref{g0})  gives a constant, system-independent number  of the molecules within each block: $g_0 ={64 \; y^3 \over (1+y)^3}$. This in turn leads to a material independent value of the average ultrasonic attenuation coefficient $\langle Q^{-1}\rangle$. The above along with  the definition given in eq.(\ref{Q0}) further suggests that the universality is brought about by the phonons of wavelength $\lambda \sim g_0 \; l$ with $l$ as their mean free path.


Taking typical value $R_m \sim 3 \; \AA$ gives $R_0 \sim 15 \; \AA$ which corresponds to the length scale for medium range topological order (MRO) ($10 \; \AA \to 30 \; \AA$). This is as expected because VWD interactions are negligible beyond MRO and other interactions start dominating beyond this length scale.

Eq.(\ref{qth}) is the central result of this paper. As described above, it is based on a balancing of the VW forces with phonon induced interactions among  molecules at MRO length scales in amorphous systems. The universal aspects of this competition, as described above, then result in the qualitative universality of $\langle  Q_{a}^{-1} \rangle_{e, \omega}$ which is consistent with experimental observations \cite{pohl}. Note, based on the type of the experiment,  the observed data for a glass often vary  from one experiment to another (see for example, the values of tunnelling strengths $C_{l,t}$ in \cite{bm,pohl}).

\section{Comparison with experimental data}

Eq.(\ref{b4}) and eq.(\ref{qth}) both give theoretical formulations for the internal friction in terms of the molecular properties. Eq.(\ref{qth}) however is based on an additional prediction $R_0= 4 R_v$, derived and analyzed  in \cite{bb3}. This motivates us to compare both predictions, namely, eq.(\ref{qth}) and eq.(\ref{b4}), with experimental data for 18 glasses given by two different studies  \cite{pohl} and \cite{bm}. 

A comparison of eq.(\ref{qth}) with experiments requires the information only about $c_l, c_t$ and is straightforward. But eq.(\ref{b4}) depends on many other material properties and needs to be rewritten as follows. As discussed in \cite{bb3}, $R_0$ can be expressed in terms of molecular properties
\begin{eqnarray}
R_0^3 = {(1+y)^6 \;  c^2 \; A_H  \; M \; \Omega_m \over    8 \;  \pi^2  \; \gamma_m^2 \; N_{av}}.
= {(1+y)^6 \;   \; A_H  \; M^2 \; c^2 \over    8 \;  \pi^2  \; N_{av}^2  \; \gamma_m^2  \; \rho_m}.
\label{r05}
\end{eqnarray}
Substitution of the relation   $\Omega_m = {4 \over 3} \pi \; R_m^3$ in eq.(\ref{r05}) gives

\begin{eqnarray}
\left(R_0\over R_v \right)^3 =
{1\over y^3} \; \left(R_0\over R_m \right)^3 &=& {(1+y)^6 \over y^3} \; {  M   \; A_H   \over    6 \;  \pi  \; N_{av} } \; \left({c  \over \gamma_m} \right)^2.
\label{r06}
\end{eqnarray}
Here $c$, as the speed of sound, and $\gamma_m$,  as the phonon mediated coupling constant between molecules, have directional dependence:  $c=c_l,c_t$ and $ \gamma_m=\gamma_l, \gamma_t$ with subscripts $l, t$ referring to longitudinal and transverse direction, respectively. 
The above along with eq.(\ref{b4}) gives, 

\begin{eqnarray}
\langle  Q_{a}^{-1} \rangle_{e, \omega} &=& {48 \; f(y) \over \eta\; \sqrt{2 \; z} \; K}  \; \left({6 \; \pi \; N_{av} \over  M \; A_H} {\gamma_a^2 \over  c_a^2} \right)^{7/4}
\label{q2}
\end{eqnarray}
where $f(y)={y^{27/4} \over (1+y)^{12}}$ with $\eta=2$, $z=3$ and the subscript $a=l,t$. For later reference,  note $f(y)$  is almost same for $y=1$ and $y=1.5$: $f(1)=2.44 \times 10^{-4}$ and $f(1.5) =2.59 \times 10^{-4}$.

As standard TTLS model is a special case of  our generic block model, the available information for the  coupling constants in the former case can be used for the latter. (Note TTLS model is based on the presence of some two level atoms/ molecules (TLS) as defects.  The  coupling constants of the molecules within a block due to molecule-phonon interaction can then be taken same as those of TLS).  The TLS coupling constants are related to tunnelling strength $C_{a}$, defined as 
\begin{eqnarray}
C_{a} = {{\overline P } \; \over \rho_m} \; \left({\gamma_a \over c_a}\right)^2, 
\label{ct}
\end{eqnarray}
 with  ${\overline P }$ as the spectral density of   tunnelling states.  According to tunnelling model, 
 \begin{eqnarray}
C_{a} = {2\over \pi} \; \langle Q_{a, pohl}^{-1} \rangle,     
\label{ct1}
\end{eqnarray}

As the experimental results are usually given in terms of TTLS parameters $C_l, C_t$, we define  the analog of $C_a$ for our case  for comparison 

\begin{eqnarray}
{\mathcal B_a} =  {2\over \pi} \; \langle Q_{a,pohl}^{-1} \rangle = 2 \; \langle Q_a^{-1} \rangle.
\label{ba}
\end{eqnarray}
The above along with eq.(\ref{q2}) and eq.(\ref{ct}) then gives
 \begin{eqnarray}
{\mathcal B}_a  &=& {6 \; f(y) \over \eta\; \sqrt{z} \; K}  \; \left({6 \; \pi \; N_{av} \over  M \; A_H} {\rho_m \; C_a \over { \overline P}}\right)^{7/4}
\label{q3}
\end{eqnarray}

\subsection{Determination of Physical Parameters} 

Both definitions in eq.(\ref{ct1}) and eq.(\ref{ba}) refer to same physical property, i.e, internal friction, thus implying $B_a = C_a$. From eq.(\ref{q3}), however, ${\mathcal B_a}$ depends on many other parameters besides $C_a$ which vary from one glass to another. Although, not obvious  a priori  how the two can be equal, this is indeed necessary if our theoretical prediction in eq.(\ref{q3}) is consistent with the experimental values for $\langle Q_a^{-1}\rangle$.
To verify the equality, we pursue a detailed quantitative analysis of ${\mathcal B}_l, {\mathcal B}_t $ for $18$ glass. The required values of $c_l, c_t$ to determine $K$ along with $\rho_m$ and $\overline{P}$ are taken from \cite{bm}. The information about $C_a, A_H$ and$M$ for the purpose is obtained as follows.

\vspace{0.1in}

{\it (i) $C_l, C_t$}: 
 Using ultrasonic absorption data, the study \cite{bm}  determines $C_l, C_t$ as adjustable parameters for $18$ glasses; these values are displayed in columns $4$ and $10$ of table II (referred as $C_{l,bm}$ and $C_{t,bm}$). The corresponding results for ${\mathcal B}_{l, t}$, derived from eq.(\ref{q3}), are displayed in columns $3$ and $9$ of table II (with notations defined in table captions).

 The $C_l, C_t$-values  mentioned in \cite{pohl} for some of the glasses are different from \cite{bm} (although   $c_l, c_t$ values are same  in both studies) which  then lead to, from eq.(\ref{ct}),  different values for $\gamma_l, \gamma_t$. Further note that \cite{pohl} considers data from two different experimental approaches, namely, acoustic  and flexural) and the results for $C_l, C_t$ values vary from one experiment to another. This motivates us to compare eq.(\ref{q3}) with two sets of data given in \cite{pohl} too. The  $C_l, C_t$ values from \cite{pohl} are displayed in Table II in columns $6,8,12,14$; the latter along with $\rho_m$ and ${\overline P}$ (both given in table I)  are used to obtain corresponding theoretical  predictions for ${\mathcal B}_l, {\mathcal B}_t$,  given in columns $5,7,11,13$.

\vspace{0.1in}

{\it (ii) $M$:}  As eq.(\ref{q3}) depends on $M^{7/4}$, a correct estimation of $M$ is important too. 
Two options available to determine $M$ are  (i) mass of the basic structural unit which dominates the structure of the glass and participates in the dispersion interaction (later referred as vwd unit), or, (ii) the molecular mass of the glass (later referred as formula unit); (here, for example for $SiO_2$ glass, 
$SiO_2$ is the formula unit but dominant structural unit can be $SiO_4$ or $Si(SiO4)$).  Clearly, with dispersion interaction as the basis of our analysis, it is reasonable to use the $1^{st}$ option . To analyze the influence however we consider both options to calculate ${\mathcal B}_l, {\mathcal B}_t$. The details of dominant structural unit for each glass and its mass, referred as $M_1$, is discussed in  {\it appendix A}.  The formula mass, labelled here as $M_2$, corresponds to weighted summation of the molar masses of each constituent of the glass: for the latter consisting of $n$  components $X_k$, $k=1 \to n$, with their molar mass as $m_k$ and weight percentage as $p_k$,  $M_2= \sum_{k=1}^n  p_k \; m_k$. The glass composition for the 18 glasses is given in {\it appendix A} and their $M_1, M_2$ values are displayed in table I.

\vspace{0.1in}

(iii)  {\it $A_H$}: for materials in which spectral optical properties are not available, two refractive-index based approximation for $A_H$ namely, standard Tabor-Winterton  approximation (TWA) (appropriate for low refractive index materials, $n < 1.8$) and single oscillator approximation (SOA) (for higher indexes  $ n > 1.8$),  provide useful estimates \cite{fr2000} . The $A_H$ for 18 glasses listed in Table 1 are based on these approximations (with details given in \cite{bb1}).

\subsection{Quantitative analysis}

As mentioned above, eq.(\ref{qth}) for $\langle Q^{-1}_{a}\rangle$ is based on relation $R_0=4 R_v$  but  eq.(\ref{q3} is based only on eq.(\ref{r03}).  The present analysis therefore provides two pathways to theoretically determine $\langle Q^{-1}_{a}\rangle$, one based on constant ratio of two short range length scales and other on molecular properties. The first pathway requires the information about $c_l, c_t$ only but the second one also requires a prior information about the tunnelling strength $C_a$. 
The reported experimental data for the latter however  varies significantly from one experiment to another (as indicated by the data in even numbered columns of tables II, III). This in turn leads to different values of  ${\mathcal B}_a$  (from eq.(\ref{q3})); the latter  are displayed in odd-numbered columns of tables II and III (for $M_1$ and $M_2$ respectively), with corresponding experimental data from \cite{bm}  and \cite{pohl} given in adjacent columns.  Note, as displayed  in table I, $M_1$ and $M_2$ do not differ significantly for the glass-ceramics and, consequently, the predictions for ${\mathcal B}_a$ for the two cases are close. However, for single component glasses e.g. SiO2 or where one component dominates (e.g. in BK7),  ${\mathcal B}_l, {\mathcal B}_t$ predictions based on $M_1$ are closer to experimental data (see table II). This in turn provides further credence to  the relevance of VW forces in present context.

The  values of ${\mathcal B}_{th}=2 \langle Q_a^{-1} \rangle$  from eq.(\ref{qth}),  along with corresponding experimental $C_a$ data for each glass, is also illustrated in figure $1$. The similar comparison based on eq.(\ref{q3}) is displayed in figure $2$ for $M=M_1$ and figure $4$ for $M=M_2$.  A direct comparison of theoretical and experimental data  is also displayed in an alternative way in figure 3 for $M_1$ and in figure 5 for $M_2$. As mentioned above, the results for a glass vary from one experiment to other often within a factor of 2 but sometimes more e.g. for polymers (see odd numbered columns of Tables II, III and also \cite{pohl}). But the deviation of our theoretical prediction from experiments is usually less than a factor of 2. 

Further, a comparison of figures $2$ and $4$ (or figures $3$ and $5$) indicates that the results for $M=M_1$ are closer to experimental data, thus indicating the molecules interacting by VWD interaction as an appropriate choice for the present analysis. This  is also consistent with our theoretical approach assuming  VWD interactions as the relevant interaction for length scales less than MRO.

An important point to note here is that the ${\mathcal B}_{a}$-dependence in eq.(\ref{q3}) on glass-properties is based only on the product $M . A_{H}$. (This can be seen by substituting $R_0 = 4 R_v$ in eq.(\ref{r05})  which  then gives the ratio ${\gamma_m\over c}$ in terms of $M. A_H$). The quantitative universality of $\langle Q^{-1}\rangle $ therefore seems to be a reconfirmation of already known relation between $A_H$ and molar volume \cite{isra}.

\section{Discussion}

The definition  in eq.(\ref{Q0}), along with an almost constant $Q^{-1}_a$, implies a linear relation between the phonon mean free path $l$ and its wavelength $\lambda$: $l \sim 10^{3} \lambda$. Within TTLS model, this behavior was explained by two different mechanisms: the  low frequency phonons were postulated to be attenuated mainly by a relaxation of TLS defects but high frequency phonons that carry the heat were believed to be resonantly scattered \cite{jack, phil2}. Later on TLS were generalized to soft local atomic potentials (quasi-harmonic oscillators) and their interactions with phonons was attributed to be the cause of a constant $Q^{-1}$ \cite{paras}. The approach however gave $C_a \sim 1$  i.e., a value three orders of magnitude too large; this later on led to suggestions that  only a small fraction of the quasi-harmonic oscillators act as tunneling defects \cite{paras, gure1}. 

Although as discussed in \cite{pohl}, TTLS model shows good agreement for many glasses,  the physical nature of tunnelling entities its not yet fully understood. Further the resemblance of the low-energy excitations in many disordered crystals to those found in amorphous solids strongly suggests their origin not related to  long-range order in materials. It is therefore necessary to seek alternative theories especially those based on MRO i.e a length scale dominated by  VW forces, present in all materials.  This is indeed the case in our approach based only on two scales $R_0$ and $R_v$, the first of the order of MRO and second that of SRO. Note ideas suggesting a role of  MRO  scales in origin of glass anomalies have appeared  in past too e.g. \cite{du,vdos,ell3,mg}. However these were based on experimentally/ numerically observed existence of structural correlations at these scales and did not explicitly consider the role of molecular interactions.

As eq.(\ref{q2}) indicates, $Q^{-1}$ depends only on the ratio $R_0 \over R_v$ which in turn is related to $g_0$, the number of molecules within the block. As the molecules interact by VW forces e.g by formation of induced dipoles that decay rapidly (i.e $r^{-6}$) with $r$ as the distance between molecules, the dominant contribution comes from the nearest neighbor molecules only. Under acoustic perturbation, the molecules go to vibrational excited state by absorbing the energy from sound waves which triggers the induced dipole interactions among neighboring molecules. As this number can not vary much from one glass to another (assuming three dimensional structure) except for thin films, this results in a constant value of $Q^{-1}$. This also explains observed deviation in some thin films (see \cite{pohl}). As indicated in table II, the value of $Q^{-1}$ given by our approach for $18$ glasses is in good agreement with experimental data.

 Further physical insight in this consistency can be given as follows. As discussed in detail in \cite{bb3}, $R_0$ is also the size of the basic block and can be expressed in terms of molecular parameters.
At large $\lambda > 2 R_0$, the basic block subunits within a macroscopic glass block respond as an array of periodic structures which in turn ensures large mean free paths, thereby reducing the attenuation. For $\lambda \le 2 R_0$ however the orientational disorder of the induced dipoles at MRO scale or less affects the phonon dynamics causing their scattering and thereby localization. Thus $R_0$ is a relevant scale for the sound absorption and thereby attenuation in glasses;
as discussed in \cite{bb3}, our $R_0$ is approximately the same as $R$ of \cite{ell3} (see table I of \cite{ell3}).
The 2nd scale $R_v$ appears in the wave-dynamics due to its sensitivity to the number of interacting  molecules  (from eq.(\ref{g0}). As the change of phonon dynamics occurs at length scale $R_0$, the Ioffe-Regel (IR) frequency $\omega_{ir}$ is therefore expected to correspond to $c_a/2R_0$, marking the transition from the well-defined acoustic like excitations to those characteristic of basic block,  with $c_a=c_l,c_t$ as the sound velocity in the medium \cite{bb3}. A comparison of our theoretical prediction $\omega_{ir}=c_a/2R_0$ with experimental available boson peak frequencies  further indicates their closeness.

At this stage, it is worth reviewing the main assumptions made to arrive at our theoretical predictions: 

(i) The  interactions  within the block are assumed to be homogeneous. The assumption was used in  section  III for the random matrix modelling of the Hamiltonian as well as in linear response theory for $Q^{-1}$. This puts an upper limit on the allowed block size.
As discussed in \cite{bb3}, the size of the block turns out to be of the medium range order  $\sim 3 nm$ with only $8$ molecules within, the assumption of homogeneity can be well satisfied. 

 Any block of bigger size would include both dispersion as well as phonon-coupling among molecules and thereby lead to inhomogeneity of the interactions. The theory in principle can still be adapted to analyze a super block consisting of bigger basic block sizes (as in \cite{vl}) but it would need many modifications including the use of sparse random matrices. (Note with a radius $R_0$, the basic block considered here satisfies this condition).

(ii)  The blocks are assumed to be of spherical shapes. This is a natural choice, keeping in view especially of the spherical shape of the molecules (although the latter is also an assumption but a standard one). It also helps a simpler technical formulation of the derivations. Alternatively, basic blocks of arbitrary shape can also be chosen but that is 
at the cost of technical complexity of intermediate steps of the derivation.  We believe that although the ratio ${R_0 \over R_m}$ may vary slightly with shape  but it will be compensated by the structure parameter $s$, thus leaving theoretical prediction in eqs.(\ref{q2},\ref{q3})  almost unaffected.

(iii) The interaction between phonon and non-phonon degrees of freedom are assumed to be weak, allowing linear response of the blocks to external perturbation.

The phonon mediated perturbation is assumed to access all $N$  levels of the basic block Hamiltonian ($N = {\mathcal N}^g =3^g$) within spectral range  $\omega_c \sim 10^{-18} \; J$ (from eq.(\ref{delb}).  Although  this gives   
the mean energy level spacing  in the spectral bulk as $\Delta_b \approx {\omega_c \over N}$ for a basic block is $\sim 10^{-22} \; J$, the mean level spacing in the lower edge of the spectrum however is much smaller and levels can be accessed by thermal perturbation at low temperatures $T \sim 1^o {\rm K}$.

(iv) The dominant interactions at at MRO length scales of the glasses are non-retarded dispersion forces among molecules. This  is applicable only to insulator glasses and needs to be replaced for other cases.

(v) The theoretical results presented here (figures 1-4 and table 1-3) are obtained from eq.(\ref{q3}) and eq.(\ref{q3}) with $y=R_v/ R_m \sim 1$ for the molecules interacting by VWD. In general $y$ fluctuates from one glass to another with $1$ as its average value; the glass-specific values for $y$ should be taken, in principle,  for better accuracy. However as noted below eq.(\ref{q2}),  $f(y)$ remains almost same for $y=1$ and $y=1.5$: $f(1)=2.44 \times 10^{-4}$ and $f(1.5) =2.59 \times 10^{-4}$.  The fluctuation of $y$ therefore does not seem to have significant effect of our results.

(vi) The ${\mathcal B}_{l}, {\mathcal B}_{t}$ values given in table II are obtained by approximate $A_H$ values used in eq.(\ref{q2}); we believe the results could be improved  if exact values of $A_H$ are used (see \cite{isra, fr2000}). Further our  results given in table II are based on the Hamaker constant of the molecules interacting in vaccum. The vwd unit is however the dominant cation surrounded by other molecules; the interaction between two cations is therefore mediated by other molecules. It is  natural to query, therefore, how the ${\mathcal B}_a$ results will be affected if $A_H$ values in the relevant medium are considered.

\section{conclusion}

\vspace{0.2in}

In the end, we summarize  with our main ideas and results.

Based on experimental evidence of ordered structure in glasses below MRO ($10 \to 30 \AA$) and its lack above, we describe a macroscopic size glass material as elastically coupled, spherical shape, generic blocks, with homogeneous dispersion  interaction within each such block. A random matrix modelling of their hamiltonian and  linear  response  to an external strain field, then  relates the low temperature averaged ultrasonic attenuation coefficient for the glass to a ratio of molecular length scales and a ratio of longitudinal and transverse sound speeds in the amorphous solid; the theoretical justification supported by numerical evidence for the former and  experimental one for the latter indicate these ratios to be almost material independent.   This in turn reveals the qualitative universality of the coefficient   which is consistent with experimental observations in the temperature regime $1^o \; K \to 10 K$ \cite{pohl}. 

The central result of our work is given by  eq.(\ref{b4}) and eq.(\ref{qth}) with main assumptions summarised in section VIII. An important insight revealed by our formulation  is the physical significance of the basic  block size: it defines a length scale $R_0$, typically of the order of MRO length scales  in glasses, beyond which $\langle Q^{-1}\rangle$ attains universal value. As discussed in \cite{bb3},  $R_0$ is the distance between two molecules at which the strength of dispersion forces between them is balanced by their phonon mediated coupling of their stress fields. Further  $R_0$ is also consistent with another assumption made in our study i.e regarding the isotropy and homogeneity of the stress filed of the basic block; this follows because  almost all molecules within a spherical block of radius $R_0$ are subjected to same interaction strength (with $8$ molecules within a basic block). 
The omnipresence of dispersion forces indicates the application of our results to other disordered materials too.

The analysis presented here takes only dispersion type inter-molecular forces into account and neglects the induction forces which restricts, in principle, the application of our results to non-polar molecules. We believe however that inclusion of induction forces would only change numerical value of $b$ ( given by eq.(\ref{b})) and would not affect the derivations given in section II-VI.  Similarly a generalization of the present theory by including electronic interactions  may explain the universality in context of metallic glasses.

\acknowledgments

I am grateful to Professor Anthony Leggett for introducing me to this rich subject and intellectual support in form of many helpful critical comments  and insights.


\newpage

\begin{sidewaystable}[ht!]
\caption{\label{tab:table I} {\bf Physical parameters for 18 glasses:}
The table lists the available data for the physical parameters appearing in eq.(\ref{b4}), eq.(\ref{q2}) and eq.(\ref{q3}).
The $\rho, c_l, c_t, {\overline P}$ data from \cite{bm} (or \cite{pohl} if not available in \cite{bm}) is displayed 
in columns $3^{rd}, 4^{th}, 5^{th}$ and $8^{th}$, respectively. 
 The columns $6^{th}$ and $7^{th}$ give the 
$\gamma_l$ and the $\gamma_t$ values, taken from  \cite{bm}  except for few cases; for those marked by a star (*), the  values are obtained either from \cite{pohl} or from $C_l, C_t$ values given in \cite{bm} along with eq.(\ref{ct1}. (Although not used for our analysis, 
the $\gamma$ values are included here for completeness).
The $A_H$ values  given in columns $9^{th}$ are taken from \cite{bb1}.
The  molar mass values, referred as $M_1$  for the vwd unit along with its composition is given in columns $10^{th}$ and $11^{th}$ and the mass $M_2$ for formula unit (same as glass molecular weight) in column $12^{th}$ respectively. }
\begin{center}
\begin{ruledtabular}
\begin{tabular}{lccccccccccr}
Index & Glass & $\rho_m$  &  $c_l$ & $c_t$  & $\gamma_l$ &  $\gamma_t$   & ${\overline P}$  &  $A_H$  & $M_1$   & Vwd unit & $M_2$ \\
\hline
 &   & $\times 10^3 {Kg/m^3}$ &  ${km/sec}$ & ${km/sec}$ & $ev$ & $ev$ & $10^{45}/J. m^3$ & $\times  10^{-20} \; J$ & $gm/ mole$  & &  $gm/mole$ \\
\hline
  1 &  a-SiO2          &    2.20 &    5.80 &    3.80 &    1.04 &    0.65  &      0.8 &  6.31  &  120.09  & [$Si(SiO_4$] &      60.08 \\
    2 &  BK7             &    2.51 &    6.20 &    3.80 &       0.9 &    0.65 &    1.1  &    7.40 &   92.81 & [$SiO_4$]       &        65.84 \\
    3 &  As2S3         &    3.20 &    2.70 &    1.46 &     0.26 &    0.17 &    2.0 &    19.07 &   32.10  & [$S$]            &     246.03 \\
    4 &  LASF             &    5.79 &    5.64 &    3.60 &    1.46 &    0.92 &    0.4 &    12.65 &167.95   & [$LASF$]      &     221.30 \\
    5 &  SF4               &    4.78 &    3.78 &    2.24 &     0.72 &    0.48 &    1.1 &    8.40  &  136.17 & [$Si_2O_5$] &    116.78 \\
    6 &  SF59            &    6.26 &    3.32 &    1.92 &     0.77 &    0.49 &    1.0 &    14.05  &   92.81 & [$SiO_4$]  &     158.34 \\
    7 &  V52               &    4.80 &    4.15 &    2.25 &     0.87 &    0.52 &    1.7 &    8.37   &  167.21 & [$ZrF_4$] &     182.28 \\
    8 &  BALNA         &    4.28 &    4.30 &    2.30 &    0.75 &    0.45 &    2.1 &        6.87  &  167.21 & [$ZrF_4$]  &      140.79 \\
    9 &  LAT               &    5.25 &    4.78 &    2.80 &    1.13 &    0.65 &    1.4 &     9.16    &  205.21 & [$ZrF_6$] &     215.69 \\
   10 &  a-Se             &    4.30 &    2.00 &    1.05 &    0.25 &    0.14  &    2.0 &   18.23 &   78.96 & [$Se$]          &     78.96 \\
   11 &  Se75Ge25  &    4.35 &    0.00 &    1.24 &               &    0.15 &    1.0 &    22.19 &  77.38  & [$Se_3Ge_1$] &     77.38 \\
   12 &  Se60Ge40    &    4.25 &    2.40*&  1.44* &            &    0.16 &    0.4 &      23.56 &76.43 & [$Se_3Ge_1$] &       76.43 \\
   13 &  LiCl:7H2O    &    1.20 &    4.00 &  2.00* &      0.62 &    0.39 &    1.4 &    4.75  &  131.32 & [$Li(H_2O)Cl_3$] &168.50 \\
   14 &  Zn-Glass      &    4.24 &    4.60 &    2.30 &    0.70 &    0.38 &    2.2 &      7.71   &  103.41 & [$ZnF_2$] &    103.41 \\
   15 &  PMMA          &    1.18 &    3.15 &    1.57 &      0.39 &    0.27 &    0.6 &      6.10 &  102.78 & [$PMMA$]  &    102.78 \\
   16 &  PS                &    1.05 &    2.80 &    1.50 &    0.20 &     0.13  &    2.8 &    6.03 &   27.00  & [$CH-CH2$] &    105.15 \\
   17 &  PC                &    1.20 &    2.97 &   1.37* & 0.28 &    0.18  &    0.9 &      6.00    & 77.10  & [$C_6H_5$]  &     252.24 \\
   18 &  ET1000        &    1.20 &    3.25 &              &    0.35 &    0.22 &    1.1 &      4.91 &   77.10   & [$C_6H_5$]   &  77.10 \\
\hline
\end{tabular}
\end{ruledtabular}
\end{center}
\end{sidewaystable}
\newpage

\begin{table}[ht!]
\caption{\label{tab:table II} {\bf Comparison of  theoretical and experimental values of internal friction for 18 glasses with $M=M_1$:} Here the theoretcial result from eq.(\ref{q3}) labelled as   ${\mathcal B}_{a,xx}$, with $a \equiv l, t$   are displayed in odd numbered columns for $M=M_1$. The $2nd$ subscript $xx$ refers to the particular experiment used to obtain required parameters in eq.(\ref{q3}): $xx \equiv bm,p1, p2$ for data from \cite{bm}, $xx \equiv p1$ for accoustic data from \cite{pohl}, $xx \equiv p2$ for flexural  data from \cite{pohl}). 
The  values used for $M_1, c_l, c_t$ to obtain ${\mathcal B}_{a,xx}$ are given in  Table I, with experimental data for $C_a$ given in  adjacent even-numbered columns.  The last column gives our theoretical prediction from eq.(\ref{qth}).}
  \begin{center}
\begin{ruledtabular}
\begin{tabular}{lcccccccccccccr}
Index & Glass  & ${\mathcal B}_{l,bm}$ & $C_{l,bm}$ & ${\mathcal B}_{l,p1}$ & $C_{l,p1}$ & ${\mathcal B}_{l,p2}$ & $C_{l,p2}$ & ${\mathcal B}_{t,bm}$ & $C_{t,bm}$ & ${\mathcal B}_{t,p1}$ & $C_{t,p1}$ & ${\mathcal B}_{t,p2}$ & $C_{t,p2}$ & ${\mathcal B}_{th}$\\
\hline
Units   &    & $\times 10^4$ & $\times 10^4$ & $\times 10^4$ & $\times 10^4$ & $\times 10^4$ & $\times 10^4$ & $\times 10^4$ & $\times 10^4$ & $\times 10^4$ & $\times 10^4$ & $\times 10^4$ & $\times 10^4$ & $\times 10^4$ \\
\hline
 1 &  a-SiO2          &    4.50 &    3.10 &    4.51 &    3.00 &    4.00 &    2.80 &    3.81 &    2.90 &    4.51 &    3.00 &    4.78 &    3.10  & 3.11\\
 2 &  BK7             &    3.09 &    2.70 &         &         &         &         &    4.38 &    3.30 &         &         &         &     & 3.01   \\
 3 &  As2S3           &    0.76 &    1.60 &    1.64 &    2.30 &    0.69 &    1.40 &    1.48 &    2.00 &    0.96 &    1.70 &         &     & 2.88   \\
 4 &  LASF7              &    1.92 &    1.20 &    4.81 &    2.00 &         &         &    1.84 &    1.16 &         &         &         &     & 3.07   \\
 5 &  SF4             &    2.58 &    2.20 &         &         &         &         &    3.89 &    2.80 &         &         &         &       & 2.97 \\
 6 &  SF59            &    4.56 &    2.30 &         &         &         &         &    6.38 &    2.80 &         &         &         &      & 2.95  \\
 7 &  V52             &    2.46 &    4.00 &    5.03 &    6.00 &         &         &    3.46 &    4.90 &    4.18 &    5.40 &         &   & 2.88     \\
 8 &  BALNA           &    1.82 &    3.80 &         &         &         &         &    2.71 &    4.80 &         &         &         &     & 2.87   \\
 9 &  LAT             &    2.29 &    3.80 &         &         &         &         &    2.15 &    3.70 &         &         &         &      & 2.96  \\
10 &  a-Se            &    0.65 &    1.20 &    0.88 &    2.20 &         &         &    0.82 &    2.20 &    1.42 &    2.90 &         &      & 2.86  \\
11 &  Se75Ge25        &         &         &         &         &         &         &         &    0.90 &         &         &      &        & \\   
12 &  Se60Ge40        &    1.86 &         &    1.83 &    1.30 &         &         &    0.14 &    0.30 &         &         &         &   & 2.99        \\
13 &  LiCl:7H2O       &    3.44 &    7.20 &    3.29 &    7.00 &         &         &    7.67 &   11.36 &    6.14 &   10.0     &         &   & 2.82        \\
14 &  Zn-Glass              &    2.09 &    3.00 &         &         &         &         &    2.79 &    3.60 &         &         &         &       & 2.82 \\
15 &  PMMA            &    1.55 &    2.00 &    4.57 &    3.70 &    3.35 &    3.10 &    4.90 &    3.70 &    7.21 &    4.80 &    9.73 &    5.70 & 2.82\\
16 &  PS              &    2.44 &    3.60 &         &         &   11.13 &    8.30 &    4.79 &    5.00 &   16.52 &   10.40 &    9.99 &    7.80   &  2.87\\
17 &  PC              &    1.00 &    1.80 &    3.51 &    3.50 &         &         &    3.19 &    3.30 &   31.23 &   12.20 &   20.16 &    9.50  & 2.77\\
18 &  ET1000    &    2.06 &    2.80 &    5.96 &    5.00 &         &         &     Inf &         &         &         &         &      & 2.52  \\  
\hline
\end{tabular}
\end{ruledtabular}
\end{center}
\end{table}

\begin{table}[ht!]
\caption{\label{tab:table III} {\bf Comparison of  theoretical and experimental values of internal friction for 18 glasses with $M=M_2$:} All other  details here are same as in table II.}
  \begin{center}
\begin{ruledtabular}
\begin{tabular}{lcccccccccccccr}
Index & Glass  & ${\mathcal B}_{l,bm}$ & $C_{l,bm}$ & ${\mathcal B}_{l,p1}$ & $C_{l,p1}$ & ${\mathcal B}_{l,p2}$ & $C_{l,p2}$ & ${\mathcal B}_{t1,bm}$ & $C_{t,bm}$ & ${\mathcal B}_{t,p1}$ & $C_{t,p1}$ & ${\mathcal B}_{t,p2}$ & $C_{t,p2}$ & ${\mathcal B}_{th}$\\
\hline
Units   &    & $\times 10^4$ & $\times 10^4$ & $\times 10^4$ & $\times 10^4$ & $\times 10^4$ & $\times 10^4$ & $\times 10^4$ & $\times 10^4$ & $\times 10^4$ & $\times 10^4$ & $\times 10^4$ & $\times 10^4$  & $\times 10^4$\\
\hline
  1 &  a-SiO2          &   15.11 &    3.10 &   15.17 &    3.00 &   13.44 &    2.80 &   12.81 &    2.90 &   15.17 &    3.00 &   16.06 &    3.10 & 3.11 \\
 2 &  BK7             &    5.64 &    2.70 &         &         &         &         &    8.00 &    3.30 &         &         &         &  & 3.01      \\
 3 &  As2S3           &    0.02 &    1.60 &    0.05 &    2.30 &    0.02 &    1.40 &    0.04 &    2.00 &    0.03 &    1.70 &         &   & 2.88     \\
 4 &  LASF7              &    1.19 &    1.20 &    2.97 &    2.00 &         &         &    1.14 &    1.16 &         &         &         &   & 3.07     \\
 5 &  SF4             &    3.37 &    2.20 &         &         &         &         &    5.09 &    2.80 &         &         &         &   & 2.97     \\
 6 &  SF59            &    1.79 &    2.30 &         &         &         &         &    2.50 &    2.80 &         &         &         &     & 2.95   \\
 7 &  V52             &    2.11 &    4.00 &    4.33 &    6.00 &         &         &    2.97 &    4.90 &    3.60 &    5.40 &         &    & 2.88    \\
 8 &  BALNA           &    2.45 &    3.80 &         &         &         &         &    3.67 &    4.80 &         &         &         &    & 2.87    \\
 9 &  LAT             &    2.10 &    3.80 &         &         &         &         &    1.97 &    3.70 &         &         &         &     & 2.96   \\
10 &  a-Se            &    0.65 &    1.20 &    0.88 &    2.20 &         &         &    0.82 &    2.20 &    1.42 &    2.90 &         &    & 2.86    \\
11 &  Se75Ge25        &         &         &         &         &         &         &         &    0.90 &         &         &         &    &     \\
12 &  Se60Ge40        &    1.86 &         &    1.83 &    1.30 &         &         &    0.14 &    0.30 &         &         &         &  & 2.99       \\
13 &  LiCl:7H2O       &    2.22 &    7.20 &    2.13 &    7.00 &         &         &    4.96 &   11.36 &    3.97 &   10.0     &         &   & 2.82     \\
14 &  Zn-Glass             &    1.35 &    3.00 &         &         &         &         &    1.80 &    3.60 &         &         &         &    & 2.77     \\
15 &  PMMA            &    1.55 &    2.00 &    4.57 &    3.70 &    3.35 &    3.10 &    4.90 &    3.70 &    7.21 &    4.80 &    9.73 &    5.70. &  2.82\\
16 &  PS              &    0.23 &    3.60 &         &         &    1.03 &    8.30 &    0.44 &    5.00 &    1.53 &   10.40 &    0.93 &    7.80  & 2.87 \\
17 &  PC              &    0.13 &    1.80 &    0.44 &    3.50 &         &         &    0.40 &    3.30 &    3.92 &   12.20 &    2.53 &    9.50 &  2.77\\
18 &  ET1000     &    2.06 &    2.80 &    5.96 &    5.00 &         &         &         &         &         &         &         &      & 2.52  \\
\hline
\end{tabular}
\end{ruledtabular}
\end{center}
\end{table}


\vfill\eject


\appendix

\section{Structural units of the glassed used for $M$ in Table 2}


The basic structural unit in a glass depends on the presence of various cations some of which act as network formers and others act as network modifiers (e.g. see pages 9-11 of \cite{b2008}). Here we give the the glass composition  and the dominant structural units for 18 glasses  used in tables I. The molar mass $M_1$ in Table I refers to  the masses of these units.

(1) $a-SiO_2$:  The 3-d network in this case has the  basic structural unit is Si[SiO4] 
with mass $M_1=120.09$ (see page 37 of \cite{b1994}, section 11.4.1 of \cite{b2011}), also see section 2.2 and fig2.7(a) of \cite{bk2011}).  

(2) BK7: (wt $\%$): The constituents in this case are $69.9 \; SiO_2$, $9.9 B_2O_3$, $8.4 NaO_2$, $8.4 K_2O$, $2.5 BaO$.  
with $70\%$ $SiO_2$, the basic structural unit in this case is [SiO4] \cite{b1978} (also see section 11.4.1, 12.1 of \cite{b2011}) with 
mass $M_1=$. 

(3) $As_2 S_3$: The glass in this case forms chain like structure e.g. $S-S$, $As-As$ or $As-S$   (i.e $S$ or $As$ of one chain interacting with neighboring one). As $S$ is dielectric, we use it as the basic unit participating in VWD interaction (see page 56 of \cite{popo}, page 125 of \cite{bori}, also see section 2.2 and fig 2.7(b) of \cite{bk2011}) and therefore choose $M_1=32$. Using molar weight of $As$  for the purpose, gives ${\mathcal B}_l=0.29, {\mathcal B}_t=0.62$ and ${\gamma_l \over c_l}= {\gamma_t \over c_t} =0.17$.

(4) LaSF7:  Also known as dense lanthanum flint glass, it  contains mostly $B_2O_3$, $La_2O_3$ and $ThO_2$ with a few $\%$ of $Ta_2O_3$ and $Nb_2 O_3$: Here the first three are main net-forming components and last two are net modifiers As net formers are in equal proportion (with each of $30\%$ weight-fraction), each one can play the role of structural unit. In this case, the structural units of each component is triangular i.e $BO_3, LaO_3, ThO_2$. The mass $M_1$ in this case is then obtained  as follows: $M = {30\over 100} (BO_3 + LaO_3 + ThO_2)$.   

(5) SF4:  The glass composition in this case is $60.8 SiO_2$, $34.9 PbO$, $2.5 K_2O$ and $1.8$ other.
with $61\%$ $SiO_2$ but with $35\%$ $PbO$, this has a basic structural unit is $Si_2O_5$ with its mass $M_1=$ 
(due to compound of type $2SiO_2. PbO$, 2 Si atoms get coordinated with $5 O$, 
(see page 17 of  \cite{b1991}, page 14-15 of \cite{b1989}).

(6) SF59: Here the constituents are $35.3.8 SiO_2$, $55.6 PbO$, $0.8 K_2O$, $7.9 B_2O_3$ and $0.4$ other material.
with reduced fraction of $SiO_2$,  the compund is of form $2PbO.SiO_2$ leading to a basic structural unit of type $SiO_4$ 
tetrahedral with $M_1=$ (see page 17 of \cite{b1991}, page 14-15 of \cite{b1989}).

(7) V52:  The constituents are $57.8 ZrF_4$, $33.8 BaF_2$ and $8.5 ThF_4$. Due to higher content of $ZrF_4$, 
the main structural unit in this case is $ZrF_4$ tetrahedral with main role of cations $Ba$ and $Th$  is to cause 2-d structure 
(network modifiers) (page 35 of \cite{b1990}, page 150 of \cite{}, page 157 of \cite{b1989}). Thus $M_1$ in this case is 
used as the molar mass of 
$ZrF_4$.

(8) BALNA: 52 $ZrF_4$, 24 $BaF_2$, 4 $AlF_3$ and  20 $NaF$. Due to higher content of $ZrF_4$, the main structural unit in this case is $ZrF_4$ tetrahedral with main role of the cations $Ba, Al$ and $Na$  is that of network modifier (i.e to cause 2-d structure) (page 35 of \cite{b1978}, page 150 of \cite{}, page 157 of \cite{b1990}). Thus $M$ in this case is used as $ZrF_4$.

(9) LAT: 60 $ZrF_4$, 33 $ThF_4$, 7 $LaF_3$. Due to higher content of $ZrF_4$, the main structural unit in this case is $ZrF_4$ tetrahedral with main role of the cations $Ba, Al$ and $Na$  is that of network modifier (i.e to cause 2-d structure) (page 35 of \cite{b1990}, page 150 of \cite{}, page 157 of \cite{b1989}). Thus $M$ in this case is used as $ZrF_6$.

(10) $a-Se$:  Se atoms  form chains or 8 atom rings through covalent/ionic bonding. The atoms on neighboring chains or rings interact by lone-pair electrons (VWD). So $M$ is that of Se atom
 (page 43 of \cite{popo}, page 115 of \cite{bori}). 

(11) $Se_{75} Ge_{25}$:  As both $Ge$ and $Se$ are network-formers, we use both atoms to calculate $M_1$ (page 115 of \cite{bori}). Thus  $M_1= M_2$ 

(12) $Se_{60} Ge_{40}$: Here again both $Ge$ and $Se$ act as network-formers (page 115 of \cite{bori}), $M_1$ is therefore obtained form both of them and therefore  $M_1=M_2$.

(13) $LiCl:7H_2O$:  The $LiCl$ molecule in presence of $H_2O$ forms $Li(H_2O)Cl_3$ tetrahedral which seems to act 
as a basic structural unit. Thus $M_1$ used in our analysis corresponds to this unit .

(14) Zn-Glass: This glass consists of $60 ZnF2$, $20 BaF_2$, $20 NaPo_3$. Due to higher content of $ZnF_2$, 
the main structural unit in this case is $ZnF_2$ with main role of the cations $Ba$ and $Na$  is that of network 
modifier (i.e to cause 2-d structure) \cite{ag}. Thus $M_1$ for this case is used as $ZnF_2$.

(15) PMMA:  The monomer here has a phenyl group $C_6H_5$ which appears as a side unit while the units along the 
main chain strongly connected by covalent bonds. As VWD interaction occurs between molecules on different chains,  
the main unit playing role here is $C_6H_5$. So $M_1$ taken is that of $C_6H_5$.

(16) PS: The monomer here has a phenyl group $C_6H_5$ as a side unit as well as $CH=CH_2$ unit while the units along 
the main chain strongly connected by covalent bonds. As VWD interaction occurs between molecules on different chains,  
the main unit playing role here seems to be  $CH-CH$ or $CH=CH_2$. The  former  could be part of Phenyl group.  Note 
unlike other polymers, the monomer of $PS$ is small and therefore 
only part of Phenyl group may be loosely held and participate in VWD.   

(17) PC: as the  monomer here is a big molecule, the Phenyl group may be loosely held and participate in VWD. 
So $M_1$ taken is that of $C_6H_5$

(18) ET1000:  here again  the  monomer is a big molecule, the Phenyl group may be loosely held and participate in VWD. So here again $M_1$ taken is that of $C_6H_5$.



\section{Relation between $\gamma$ and $\gamma_m$ }

Consider the linear response of a basic block, labeled as $''s''$ containing $g_0$ molecules, to an external strain field. The existence of long wavelength  phonons at low temperatures leads to a phonon-mediated pair-wise interaction among molecules, decaying as inverse cube of distance between them.
Consider two molecules, labeled as  ''1'' and  ''2'' with their centers at a distance $r$ within the block.  Following the same formulation as in case of blocks,  and with $T_{\alpha \beta}$ as the stress tensor component for the molecule, the corresponding  interaction energy can be written as
\begin{eqnarray}
V_{stress}({\bf r})  &=& {1\over \alpha_0 \pi \rho_m c^2}   \; \sum_{te} \;  
{\kappa^{(12)}_{\alpha \beta \gamma \delta}  \over   | \; {\bf r}_1-{\bf r}_2 \; |^3 } \; \;   T^{(1)}_{\alpha \beta} \otimes\; T^{(2)}_{\gamma \delta} 
\label{mo1}
\end{eqnarray}
with $\alpha_0=4$, $r=|{\bf r}_1-{\bf r}_2|$, $\kappa^{(12)}_{\alpha \beta \gamma \delta}  $  as in the case of  block-block interaction (given by eq.(\ref{ad3})), $\rho_m$ as the mass-density of the molecule, $c$ as speed of the sound waves. The ensemble averaged interaction energy can then be approximated as 
\begin{eqnarray}
V_{stress}(r) \approx  {\gamma_m^2 \over \rho_m \; c^2 \; r^3}
\label{mo2}
\end{eqnarray}
with  $\gamma_m$ as  the average strength of the phonon induced $r^{-3}$ coupling of  the two molecules.

The interaction parameter $\gamma_m$ can be determined as follows.
As ${\rm Tr} (V_{stress}) =0 $, one can write, with $\langle . \rangle_e$ as the ensemble average
\begin{eqnarray}
\langle {\rm Tr}  (V_{stress}^2) \rangle_e &\approx&  \left({\gamma_m^2 \over \rho_m \; c^2 \; r^3} \right)^2     
\label{mo3}
\end{eqnarray}
But
\begin{eqnarray}
\langle {\rm Tr}  (V_{stress}^2) \rangle_e 
&\approx & \left({1\over \alpha_0 \pi \rho_m c^2 \; r^3}\right)^2   \; \sum_{te, te'} \;  
\kappa^{(12)}_{\alpha \beta \gamma \delta}  \;\kappa^{(12)}_{\alpha' \beta' \gamma' \delta'}  \;  \langle {\rm Tr} \left( T^{(1)}_{\alpha \beta} \; T^{(2)}_{\gamma \delta} T^{(1)}_{\alpha' \beta'} \; T^{(2)}_{\gamma' \delta'} \right)\rangle_e
\label{mop3}
\end{eqnarray}
Further as
\begin{eqnarray}
{\rm Tr} \left( T^{(1)}_{\alpha \beta} \; T^{(2)}_{\gamma \delta} T^{(1)}_{\alpha' \beta'} \; T^{(2)}_{\gamma' \delta'} \right) =
\sum_{n,m,k,l} T^{(1)}_{\alpha \beta;nm} \; T^{(2)}_{\gamma \delta;mk} T^{(1)}_{\alpha' \beta';kl} \; T^{(2)}_{\gamma' \delta';ln} 
\label{moo3}
\end{eqnarray}
with $T^{(1)}_{\alpha \beta;nm} \equiv \langle n |T^{(1)}_{\alpha \beta} | m \rangle$, state $|n \rangle$ referring to one of the ${\mathcal N}$ single molecule states (unperturbed). Following similar ideas as in the case of a block, we have  $\langle T^{(1)}_{\alpha \beta; mn} T^{(2)}_{\alpha \beta; kl} \rangle =0$  $\forall m,n,k,l$ and $\langle T^{(1)}_{\alpha \beta; mn} T^{1)}_{\alpha' \beta'; kl} \rangle =\tau^2 \; \delta_{\alpha \alpha'} \delta_{\beta \beta'}(\delta_{nk} \delta_{ml} + \delta_{nl} \delta_{mk})$. On ensemble average, the above leads to 
\begin{eqnarray}
\langle {\rm Tr} \left( T^{(1)}_{\alpha \beta} \; T^{(2)}_{\gamma \delta} T^{(1)}_{\alpha' \beta'} \; T^{(2)}_{\gamma' \delta'} \right) \rangle &=&
2 \;  \sum_{m,n} \langle \left( T^{(1)}_{\alpha \beta;nm} \right)^2 \rangle \; \langle \left(T^{(2)}_{\gamma \delta;mn}\right)^2 \rangle \; \delta_{\alpha \alpha'} \delta_{\beta \beta'} \delta_{\gamma \gamma'} \delta_{\delta \delta'} \\
&=& 2 \; {\mathcal N}^2 \; \tau^2 \; \delta_{\alpha \alpha'} \delta_{\beta \beta'} \delta_{\gamma \gamma'} \delta_{\delta \delta'}
\label{moo4}
\end{eqnarray}
The above on substitution in eq.(\ref{mo3}) leads to
\begin{eqnarray}
{\rm Tr}  (V_{stress}^2) 
&\approx & 2 \; \left({{\mathcal N} \; \tau^2 \over \alpha_0 \pi \rho_m c^2 \; r^3}\right)^2   \; \sum_{te} \;  
(\kappa^{(12)}_{\alpha \beta \gamma \delta} )^2  \;  
\label{mo4}
\end{eqnarray}
Comparison of eq.(\ref{mo4}) with eq.(\ref{mo3}) gives 
\begin{eqnarray}
\gamma_m^2 
&\approx & {{\mathcal N} \; \tau^2 \over \alpha_0 \pi }   \;   K \; \sqrt{2}
\label{mop4}
\end{eqnarray}
where $K^2 = \sum_{te} \;  (\kappa^{(12)}_{\alpha \beta \gamma \delta} )^2$.

To relate the above to basic block property $\gamma^2$, we proceed as follows. The stress-operator for a basic block can be written in terms of those of molecules:
$$\Gamma^{(s)}_{\alpha \beta; mn} = \sum_{a=1}^{g_0} T^{(a)}_{\alpha \beta; mn}.$$
The subscripts $m, n$ now refer to an arbitrary pair chosen from $N={\mathcal N}^{g_0}$ many body states of the basic block (e.g.  the product states $|e^0_m\rangle$ and $|e^0_n\rangle$ of single molecule states).  Further $T^{(x)}_{\alpha \beta; mn} \not= 0$ only if $|e^0_m\rangle$ and $|e^0_n\rangle$ differ only by the contribution from the $x^{th}$ molecule; this leaves only ${\mathcal N}^{g_0+1}$ non-zero matrix elements for each $T^{(x)}$. Noting that stress marix elements of different molecules are uncorrelated, 
it can now be shown that 
\begin{eqnarray}
\sum_{m, n=1}^{N} \langle
\Gamma^{(s)}_{\alpha \beta; mn}  \;  \Gamma^{(s)}_{\alpha \beta; n m } \rangle_{e}=
\sum_{m, n=1}^{N} \langle
\left(T^{(a)}_{\alpha \beta; mn}\right)^2  \rangle_{e} = 
 g_0 \; {\mathcal N}^{g_0+1}  \; \tau^2 =g_0 \; N \;  {\mathcal N}  \; \tau^2.
\label{mo5}
\end{eqnarray}
Further assuming homogeneous interaction within a basic block,  the variances of all matrix elements of the basic block can be approximated as almost equal.  The left side of eq.(\ref{mo5}) is then equal to $N^2 \; \nu^2$ (with  $\langle (\Gamma^{(s)}_{\alpha \beta; mn})^2\rangle_{e}=\nu^2$) which leads to 

\begin{eqnarray}
\gamma^2 =g_0 \;  {\mathcal N}  \; \tau^2   = {g_0 \alpha_0 \pi \over K \sqrt{2}} \;   \gamma_m^2
\label{mo6}
\end{eqnarray}

Taking  $\kappa^{(12)}_{\alpha \beta \gamma \delta}$ from eq.(\ref{ad3})), we have 
$K^2 = 18 \left[1+4 \left(1-{c_t^2 \over c_l^2}\right)\right]$ 

\newpage


\section {Abbreviations}

\vspace{0.1in}

$a_0$: Bohr's radius

$c_a$:  speed of sound in longitudinal or transverse direction

$g_0$: number of molecules in a block

${\bf r}_k$: position of the molecule labeled $''k''$

$m$: mass of the molecule

${\mathcal H}$: Hamiltonian of the block with interaction among molecules included.

${\mathcal H}_0$: Hamiltonian of the non-interacting molecules within a block

${\mathcal H}^{(n)}_0$:   Hamiltonian of a  single molecule labeled $''n''$

${\mathcal U}$: molecular interactions among molecules

${\mathcal N}$: number of roto-vibrational states in the electronic ground state of each molecule 
playing role in the analysis:  ${\mathcal N}=3$

$N ={\mathcal N}^{g_0}$:  size of the eigen-basis space of the basic block  

$\eta$: number of allowed dipole transitions among roto-vib states in the electronic ground state of a single molecule 

${\mathcal K}$: an eigenstate of the basic block which is a product state of single molecule eigenstates  

$|{\mathcal K}_n \rangle$: eigenstate of a single molecule labeled ''n'' which contributes to the basic block eigen-state ${\mathcal K}$

$\nu_0 = \langle  E_{\mathcal K_n}^2\rangle$: the variance (the square of the line-width) of the vibrational energy-levels of a single isolated molecule Hamiltonian ${\mathcal H}_0^{(n)}$.

$b$: bulk spectral parameter: ${1\over b^2} = 2 \sum_{\mathcal L} \; v_{_{\mathcal KL}}$

$R_m$:  distance between two nearest neighbor molecules

$2 R_v$:  distance of closest separation between two neighbor molecules 
(so total distance between their centers is $2(R_m+R_v) \approx 4 R_m$

$2 R_0$: distance between centers of two nearest neighbor basic blocks

$\Omega_m$:  volume of a molecule,  assumed spherical

$\Omega_{eff}$: effective volume occupied by a molecule i.e including inter-molecule separation
$\Omega_{eff}=8 \Omega_m$

$\Omega_b$: volume of a basic block

$\rho_m$: mass density of a basic block

$\rho_n$: particle density in a basic block

$c$: speed of sound waves in the glass block

$\gamma_m$: strength of the phonon-mediated $r^{-3}$ interaction  between two molecules 
($\gamma_m \equiv \gamma_l$ or $\gamma_t$ based on longitudinal or transverse direction

$\gamma^2$: $=N \nu^2$  where $\nu^2$ is the variance of  the stress matrix element of a basic block. 


$\rho(e)$: spectral density of the basic block interaction Hamiltonian ${\mathcal H}$

$A_H$: Hamaker constant in Vacuum

$C_6$:  strength of dispersion interaction 

$V_{dispersion}$: dispersion interaction between two molecules at a distance $r$: $= C_6/ r^6$

$V_{stress}$: phonon mediated interaction between two molecules

$V_{b, dispersion}$: dispersion interaction between two neighbor basic blocks

$V_{b, stress}$: phonon mediated interaction between two neighbor basic blocks

$\langle Q_{a, Pohl} \rangle$:  internal friction defined in \cite{pohl},  $a=l,t$

$\langle Q_{a, leggett} \rangle$:   internal friction defined in \cite{vl}

$C_a$:  tunneling strength defined as $C_a={\gamma_a^2 {\overline P}\over c_a^2 \; \rho_m}$ 

$C_{a,bm}$:  tunneling strength data from \cite{bm}

$C_{a, p1}$:  tunneling strength data from \cite{pohl} ( acoustic experiment) 

$C_{a, p2}$:  tunneling strength data from \cite{pohl} (flexural experiment) 

$M_1$: molar mass of the dominant unit in the glass structure

$M_2$: molar mass of the basic formula of the glass 

${\mathcal B}_a$: $={2\over \pi} \; \langle Q_{a, Pohl} \rangle$ 

${\mathcal B}_{a1}$:  reference to value obtained from eq.(\ref{q2}) by taking $M=M_1$

${\mathcal B}_{a2}$:  reference to value obtained from eq.(\ref{q2}) by taking $M=M_2$

\newpage

\begin{figure}[ht!]
	\centering
	\includegraphics[width=1.0\textwidth,height=0.8\textwidth]{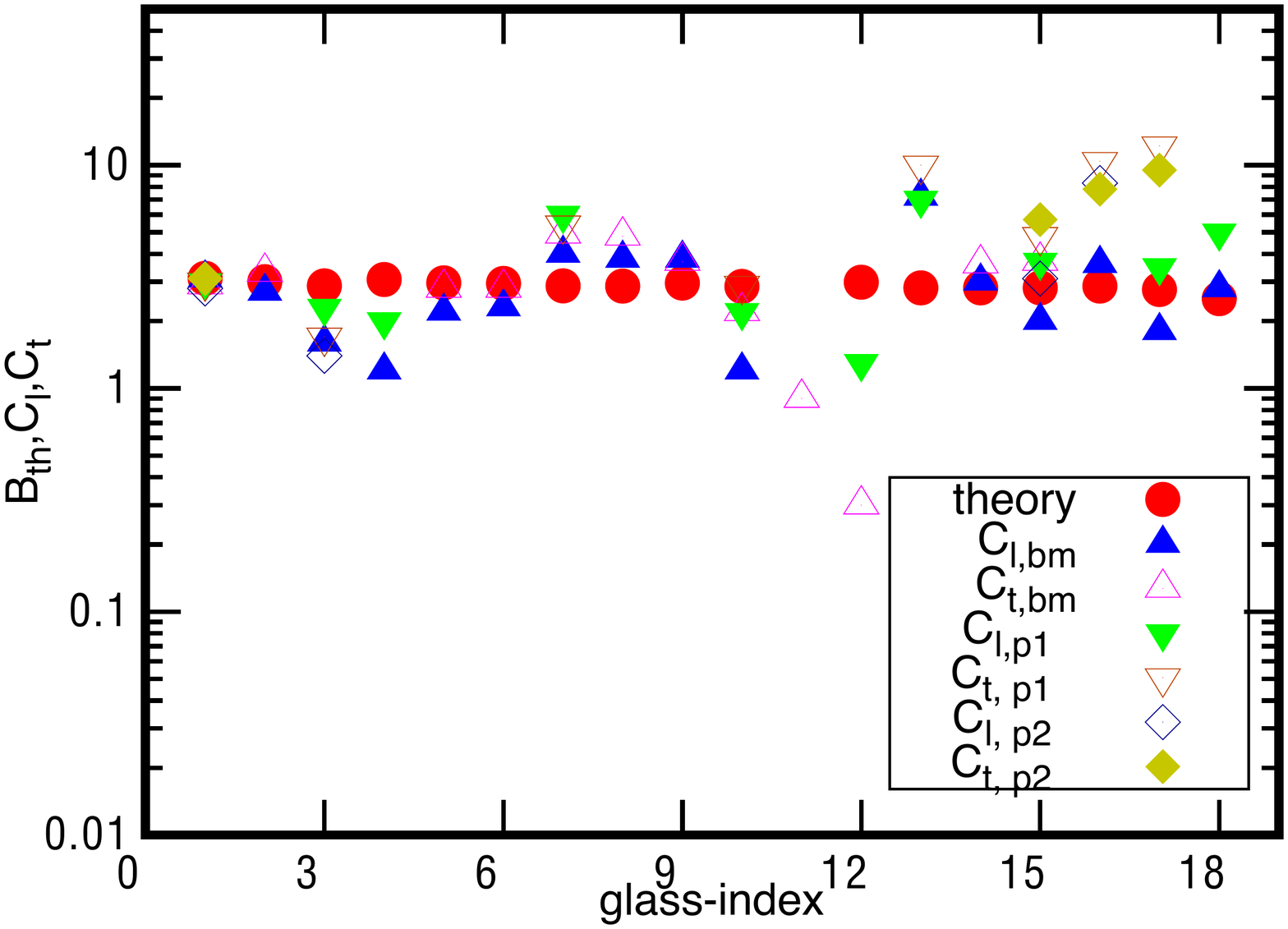}
	\caption{{\bf   ${\mathcal B}_{th}$-values for $18$ glasses}: The figure depicts the theoretically predicted ${\mathcal B}_{th}$  from eq.(\ref{qth}) and corresponding experimentally known tunneling strengths $C_{a}$ with respect to glass-index (given in $1^{st}$ column of Table I). The symbol $C_{a,bm}$  refers to experimental data for tunneling strength from \cite{bm} and  $C_{a,p1}$, $C_{a,p2}$   to  acoustic and  flexural data, respectively, from  \cite{pohl}. The values for $B_{th}$ are also given in the last column of table II and III; note these are same for both $M_1, M_2$}.
\label{fig1}
\end{figure}

\begin{figure}[ht!]
	\centering
	\includegraphics[width=1.0\textwidth,height=1.0\textwidth]{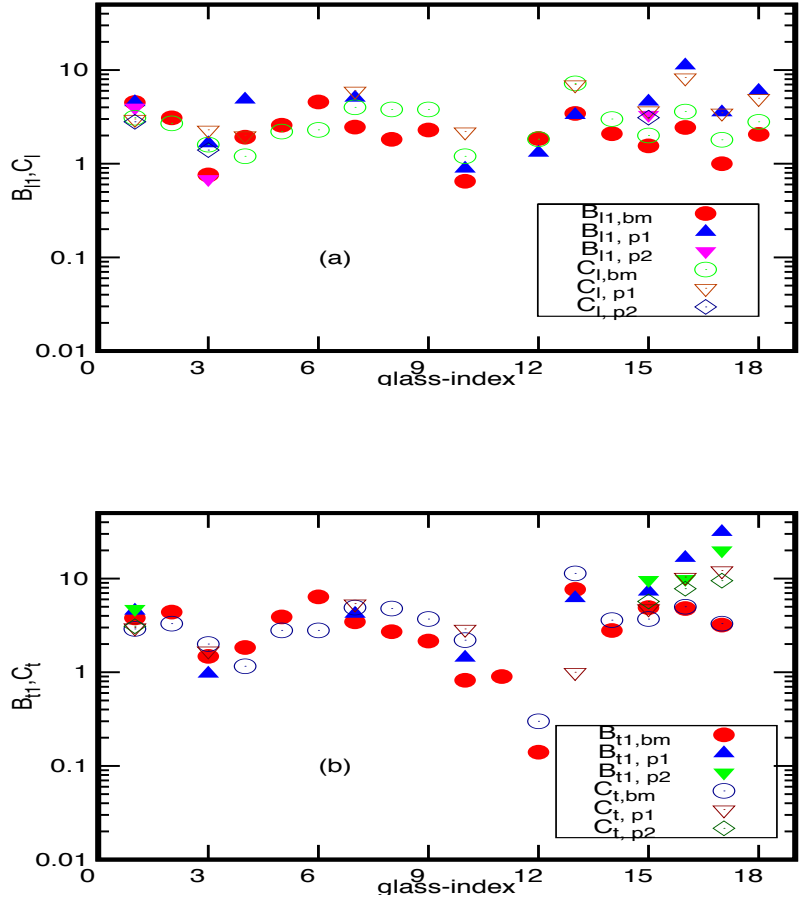}
	\caption{{\bf   ${\mathcal B}_{a}$-values for $18$ glasses (with $M=M_1$)}: The figure depicts the theoretically predicted ${\mathcal B}_{a}$  and corresponding experimentally known tunneling strengths $C_{a}$ with respect to glass-index (all listed in Table II). Here ${\mathcal B}_{a,xx}$  refers to eq.(\ref{q3}) using tunneling parameters from different experiments  (with $xx=bm$ referring to experimental data from  \cite{bm},  $xx = p1$ to  acoustic and $xx=p2$  to flexural data from \cite{pohl}). The symbols $C_{a,xx}$  refer to experimental data from \cite{bm} and \cite{pohl} accordingly. }
\label{fig2}
\end{figure}

\begin{figure}[ht!]
\vspace{-3in}
	\centering
	\includegraphics[width=1.2\textwidth,height=1.7\textwidth]{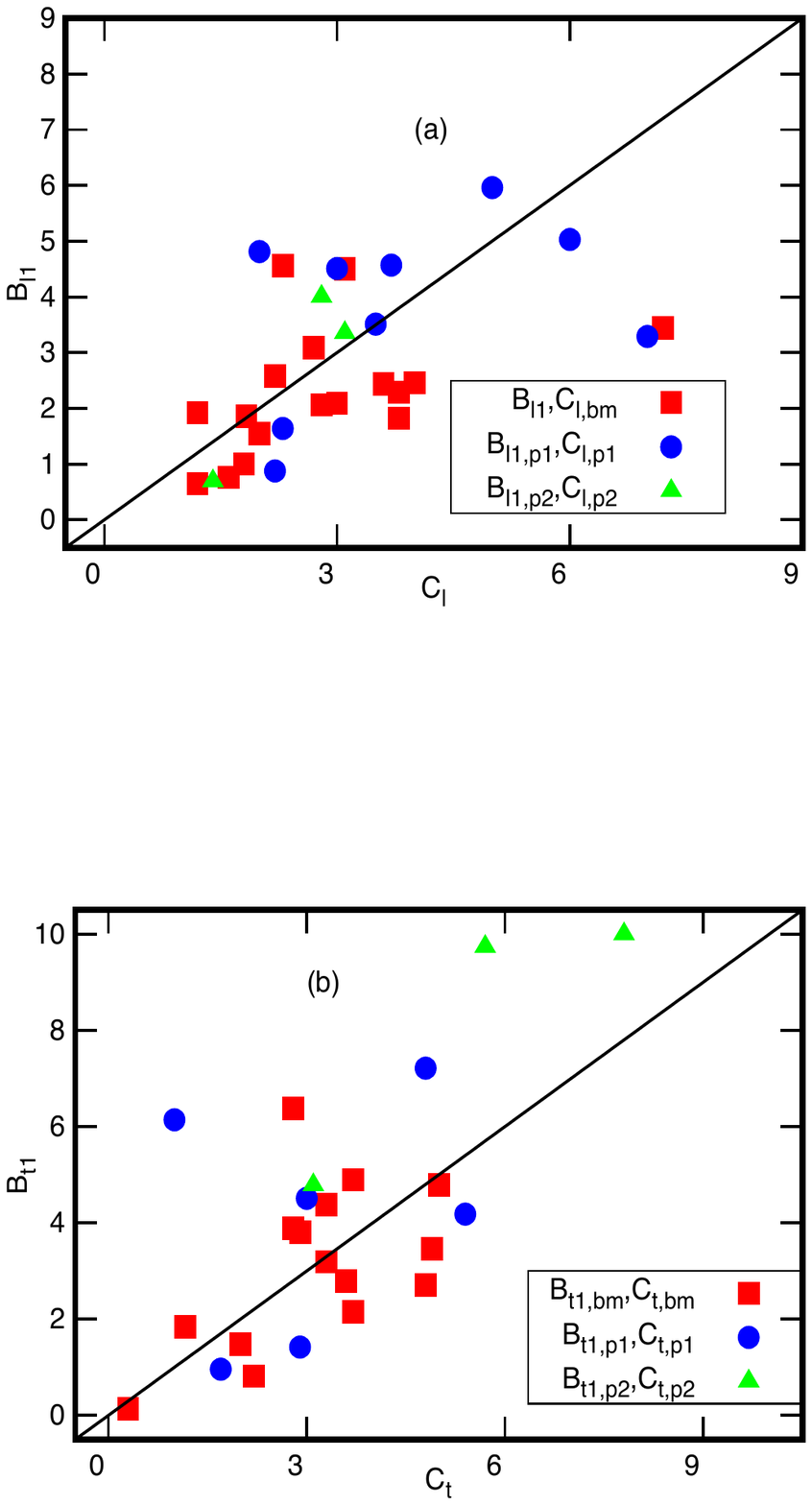}
	\caption{{\bf  Comparison of  ${\mathcal B}_a$-values ($a=l,t$),  for $18$ glasses 
	from eq.(\ref{q3}), for $M=M_1$,  with their experimentally known tunneling strengths }: here  the  ${\mathcal B}_{a1,xx}$-values correspond to $y$-coordinates of the points marked on the figure and $C_{a,xx}$ to their $x$-coordinates; the details of the labels are same as in figure 2. Here the solid line is shown only for visual guidance.} 
\label{fig3}
\end{figure}

\begin{figure}[ht!]
	\centering
	\includegraphics[width=1.0\textwidth,height=1.0\textwidth]{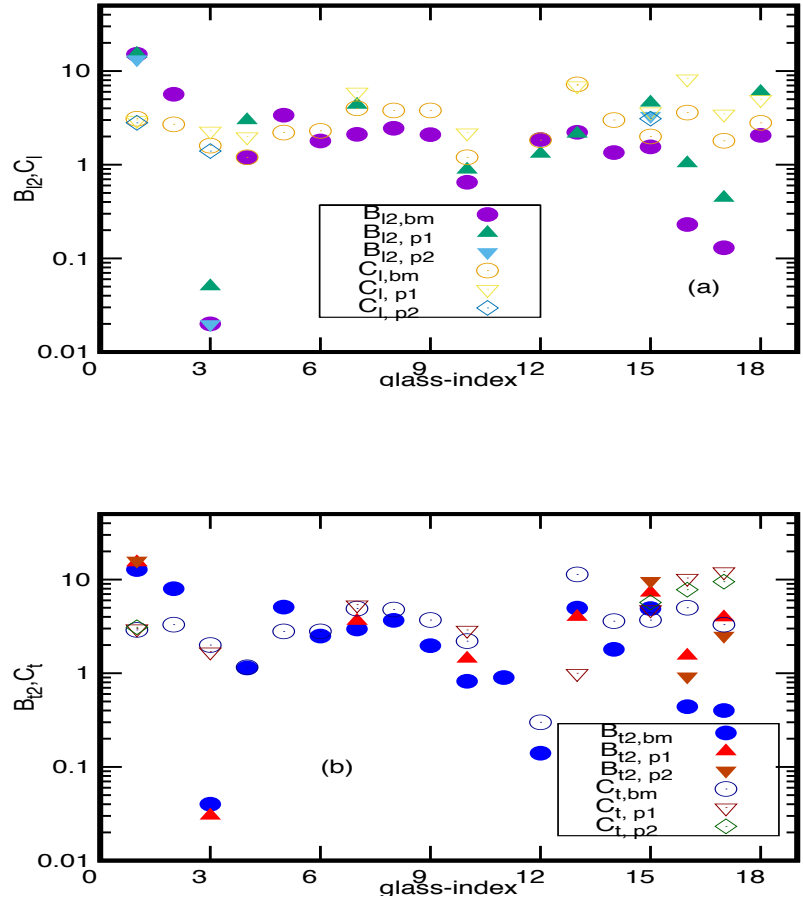}
	\caption{{\bf   ${\mathcal B}_{a}$-values for $18$ glasses}:  All details are same as in figure 1 except that now the  results for ${\mathcal B}_{a,xx}$ from eq.(\ref{q3}) correspond to $M=M_2$.  Note although the correspondence with experiments here is not as good as for $M_1$, the deviation however is still within a factor of 10. As reported in \cite{pohl}, the deviation of different experimental results lies also within that range.
}
\label{fig4}
\end{figure}

\begin{figure}[ht!]
\vspace{-3in}
	\centering
	\includegraphics[width=1.2\textwidth,height=1.7\textwidth]{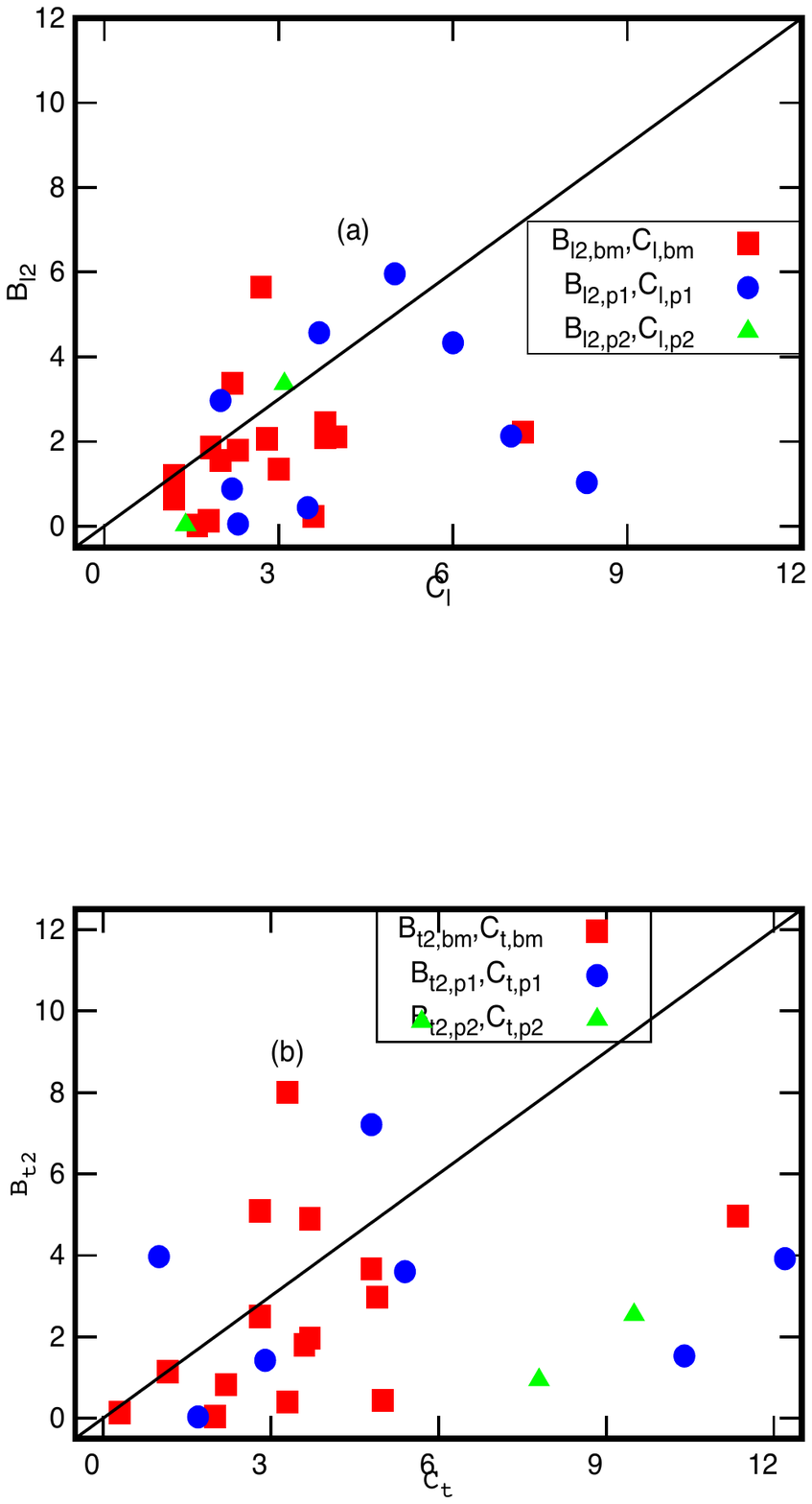}
	\caption{{\bf  Comparison of  ${\mathcal B}_a$-values ($a=l,t$),  for $18$ glasses 
	from eq.(\ref{q3}), for $M=M_2$,  with their experimentally known tunneling strengths }: here  the  ${\mathcal B}_{a1,xx}$-values correspond to $y$-coordinates of the points marked on the figure and $C_{a,xx}$ to their $x$-coordinates; the details of the labels are same as in figure 2. Here again the solid line is shown only for visual guidance.} 
\label{fig5}
\end{figure}

\end{document}